\documentclass[pre,10pt,twocolumn,superscriptaddress,floatfix,noeprint]{revtex4-1}
\usepackage{amsmath}
\usepackage{latexsym}
\usepackage{amssymb}
\usepackage{graphics,epstopdf}
\usepackage{hyperref}
\usepackage[caption=false]{subfig}
\hypersetup{
	colorlinks = true,
	linkcolor =blue,
	citecolor=blue, 
	urlcolor=blue 
}

\usepackage{mathtools}

\newcommand{\Tr}{\ensuremath{\mathrm{Tr}\,}}
\newcommand{\avg}[1]{\ensuremath{\langle #1 \rangle}}

\newcommand{\ket}[1]{\ensuremath{\vert #1 \rangle}}
\newcommand{\bra}[1]{\ensuremath{\langle #1 \vert}}

\MakeRobust{\eqref}
\newcommand{\eqnref}[1]{Eq.~\eqref{#1}}
\newcommand{\figref}[1]{Fig.~\eqref{#1}}
\newcommand{\hs}{\hat{\sigma}}

\newcommand{\hrho}{\hat{\rho}}
\newcommand{\Hop}{\hat{H}}
\newcommand{\Uop}{\hat{U}}

\newcommand{\hM}{\hat{M}}

\newcommand{\nn}{\mathbf{n}}
\newcommand{\Piop}{\hat{\Pi}}

\newcommand{\popa}{p_{\mathcal{M}}}
\newcommand{\popb}{p_{\beta}}
\newcommand{\popaum}{p_{\mathcal{M},\mathrm{um}}}
\newcommand{\popam}{p_{\mathcal{M},\mathrm{m}}}

\let\oldcdot\cdot
\usepackage{breqn}
\let\cdot\oldcdot

\makeatletter
\let\cat@comma@active\@empty
\makeatother
\begin{document}
\title{Two-stroke Quantum Measurement Heat Engine}
\author{M. Sahnawaz Alam}
\email{sahnawaz.alam@pwr.edu.pl}
\affiliation{Indian Institute of Technology Gandhinagar, Palaj, Gujarat 382355, India}
\affiliation{Wroclaw University of Science and Technology, Wroclaw, Poland}
\author{B. Prasanna Venkatesh}
\email{prasanna.b@iitgn.ac.in}
\affiliation{Indian Institute of Technology Gandhinagar, Palaj, Gujarat 382355, India}

\begin{abstract}
	We propose and analyze the theoretical model for a two-stroke quantum heat engine with one of the heat baths replaced by a non-selective quantum measurement. We show that the engine's invariant reference state depends on whether the cycle is monitored or unmonitored via diagnostic measurements to determine the engine's work output. We explore in detail the average work output and fluctuations of the proposed heat engine for the monitored and unmonitored cases. We also identify unitary work strokes for which the invariant states can support coherences in the energy basis leading to differing predictions for the average energy change during the unitary work strokes and the average work from the standard two-projective measurement approach.
\end{abstract}

\maketitle
\section{Introduction}

A central challenge in the study of quantum heat engines is to identify quantum effects that distinguish their operation from analogous classical machines or elucidate scenarios that are impossible to realise in a classical setting \cite{Vinjanampathy2016,Alicki2018,Levy2018,Kammerlander2016}. One way to address this challenge is to examine progressively the effect of including different aspects of quantum mechanics such as quantization of levels, coherences, quantum correlations \cite{Levy2018}, and possibility of driving engines using non-standard reservoirs such as squeezed thermal baths \cite{Rossnagel2014,Niedenzu2016} or finite sized baths \cite{Pozas2018}. An alternate strategy is to consider the thermodynamics implications of what is arguably the most unique feature of quantum mechanics - quantum measurements \cite{Jacobs2012}. The application of this strategy to quantum heat engines has led to the nascent field of studying `Quantum Measurement Engines' \cite{Yi2017,Jordan2020} which are quantum heat engines fueled by quantum measurements as opposed to conventional thermal reservoirs. Starting from the seminal contribution of \cite{Yi2017}, this field has quickly grown as an active area of research. A non-exhaustive list of previous work includes studies examining quantum Otto cycles with one of the thermal baths by a non-selective measurement of different kinds of working fuel (WF) quantum systems \cite{Ding2018a,Anka2021,Lin2021,Behzadi2021,Su2021},  engines entirely fuelled by quantum measurements \cite{Elouard2018}, using quantum measurement to fuel two-stroke absorption refrigerators \cite{Buffoni2019}, entangled two-qubit heat engines operated by measurement and feedback \cite{Bresque2021}, and work extraction from thermal states by phase sensitive homodyne measurements \cite{Opatrny2021}. 

In this paper we propose and analyze the theoretical model for a two-stroke heat engine with bipartite working fuel systems \cite{Quan2007,Campisi2015,Sacchi2021a,Sacchi2021b} with one of the thermal heat baths replaced by a non-selective quantum measurement. The measurement 'bath' will be described by a collection of Positive Operator Valued Measure (POVM) operators. Conventional two stroke heat engines involve a working fuel made up of two sub-systems $A$ and $B$ that are subject to a unitary work stroke corresponding to an interaction that is turned on and off between the sub-systems. This is followed by a heat stroke where the systems (that no longer interact with each other) are subject to individual thermal baths \cite{Quan2007,Campisi2015,Sacchi2021a,Sacchi2021b}. In this manner, they provide a clear realization of a quantum heat engine with heat exchange and work exchange separated. This can also be viewed as a set-up that is somewhat in-between non-autonomous cyclic heat engines and autonomous heat engines where a quantum system is simultaneously connected to two reservoirs of different temperature. As opposed to such conventional two-stroke heat engines, the two-stroke measurement (TSM) heat engine we present here exhibits the following novel feature - the invariant reference state that the system cycles back to is dependent not just on the measurement `bath' and the unitary work stroke but crucially on the details of the diagnostic procedure used to determine work statistics. In particular we show that the reference state depends on whether the cycle is interspersed with monitoring diagnostic measurements for the energy exchanges \cite{Ding2018a,Son2021}.

The paper is organized as follows. In Sec. \eqref{sec:sec2} we describe the set-up and of the two-stroke measurement (TSM) heat engine for general bipartite working fuel (WF) systems. Without recourse to a particular model for the system or the exact measurement, we present the central idea of the paper, namely the calculation of the invariant reference state corresponding to a given unitary work stroke and non-selective measurement with or without diagnostic measurements for the energy exchanges. In Sec. \eqref{sec:sec3} we exemplify our central idea by considering a two-qubit working fuel system. We use the two-projective measurement approach (TMA) for the diagnostic measurements. Here we consider a total of three combinations of operator and unitary operators. The first combination leads to the same average work with or without diagnostic measurement and the second combination produces different average work for the reference state with diagnostic measurements.  In the final case we adopt an unitary operator that leads to a reference state with coherence such that the resulting average work determined by TMA is not the same as the average energy change in the unitary stroke \cite{Solinas2015,Talkner2016,Perarnau-Llobet2017}. In addition to characterizing the engines using the average work output, we also examine fluctuations as characterized by the reliability (defined as the ratio of average work to its standard deviation). We conclude in Sec. \eqref{sec:sec4} with a summary of our findings, brief comments on possible experimental realization of such two-stroke measurement heat engines, and some outstanding questions arising from our work that can be examined in the future .

\section{Set-up: General Two-stroke Measurement Heat Engine}
\begin{figure}[tbh!]
	\includegraphics[width=\linewidth]{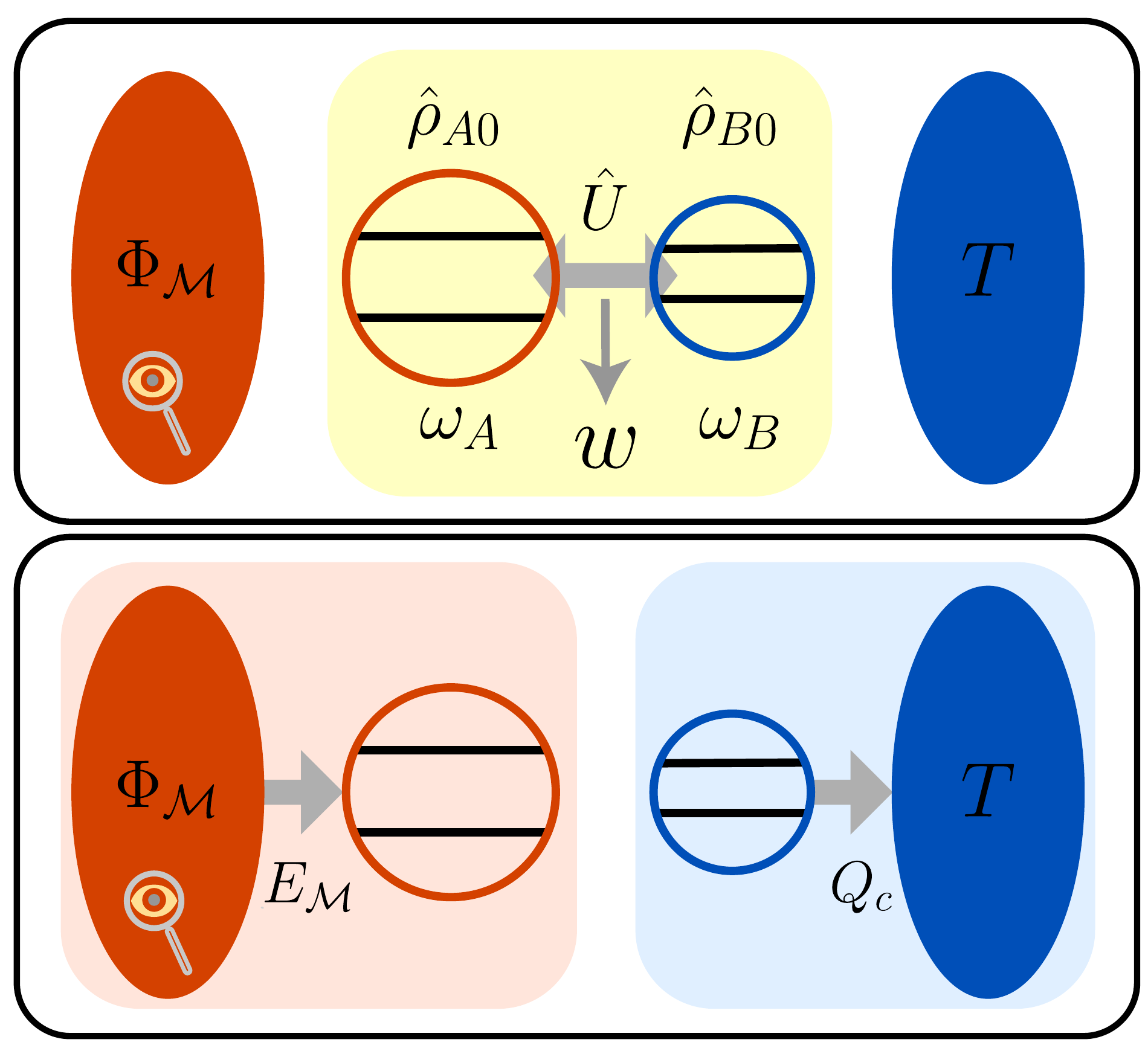}
	\caption{(Color online). Schematic representation of the two-stroke measurement heat engine. Initially the two WF systems start in a product state $\hrho_{A0} \otimes \hrho_{B0}$ with sub-system $B$ in equilibrium with the cold bath at temperature $T$. In the unitary work stroke, the two sub-systems are decoupled from the baths and a coupling unitary operation $\Uop$ is applied and leads to work extraction. In the second stroke, sub-system $A$ is subject to a non-selective measurement represented by the map $\Phi_{\mathcal{M}}$ and receives energy $E_{\mathcal{M}}$ and $B$ thermalizes with the cold bath and in the process rejecting heat $Q_c$. For the sake of definiteness, we have taken the systems as qubits with energy gap $\omega_A$ and $\omega_B$ respectively (see Sec. \eqref{sec:sec3} )} 
	\label{fig:figure1}
\end{figure}

\label{sec:sec2}
Our set-up consists of a bipartite working fuel (WF) system composed of two sub-systems $A$ and $B$ with the bare hamiltonians $\Hop_A$ and $\Hop_B$, a thermal heat bath at temperature $T = 1/(k_B\beta)$ and a measurement-based energy bath shown in \figref{fig:figure1}. Initially the WF system is in the tensor-product state with system $B$ in a Gibbs thermal equilibrium state:
\begin{align}
	\hrho_0 = \hrho_{A0} \otimes \hrho_{B0} = \hrho_{A0} \otimes \frac{e^{-\beta \Hop_B}}{Z_{B}(\beta)}, \label{eq:initstate}
\end{align}
with $Z_{B}(\beta) = \Tr_B(e^{-\beta \Hop_B})$ is the partition function. For now, we will leave the initial state $\hrho_{A0}$ of $A$ unspecified.  We first describe the engine cycle without diagnostic measurements to monitor energy exchanges. Starting from the initial state \eqnref{eq:initstate}, after system $B$ is isolated from the thermal bath, the two systems are allowed to interact with each other. This interaction is enabled by an external agent that implements a global unitary evolution given by the operator $\hat{U}$ on the total system $A+B$ actualizing the work stroke. Following this, in the second `heat'-exchange stroke, the two subsystems are uncoupled and system $B$ is brought into contact with the thermal heat bath and allowed to equilibrate with it. This thermalization step resets the state of sub-system $B$ to the initial state $\hrho_{B0}$ irrespective of the state immediately after the unitary work stroke. On the other hand the system $A$ is subject to a non-selective quantum measurement `bath' which we characterize by an operation $\Phi_{\mathcal{M}}$ i.e. given the density matrix $\hrho_A$ prior to the measurement, the post-measurement state is given by $\hrho_{\mathrm{pm}} = \Phi_{\mathcal{M}}(\hrho_A)$. Here, the post-measurement state clearly depends on the pre-measurement state. While we keep the trace-preserving operation $\Phi_{\mathcal{M}}$ general at this point, we assume that this map is completely positive and is unital. In the next section, while discussing specific realizations of the engine cycle we will write this operation in terms of minimally disturbing hermitian POVM operators. Thus, after this stroke system $B$ is restored to its initial Gibbs state (see Eq. \eqref{eq:initstate}). In contrast, sub-system $A$'s final state after passing through the two strokes of the  unmonitored cycle can be written as the following map on its initial state $\hrho_{A0}$:
\begin{align}
	\Phi_{\mathrm{um}} [\hrho_{A0}] =  \displaystyle \sum_{k=1}^{n_l} \Phi_{\mathcal{M}} \left( \Tr_B \left(  \Uop \hrho_{A0} \otimes \hrho_{B0} \Uop^{\dagger}\right) \right) \label{eq:unmonmap},
\end{align}
with $ \Tr_B$ denoting a partial trace over sub-system $B$. Note that in order that the measurement exchanges energy with the sub-system $A$, we require that the map does not leave the hamiltonian invariant \emph{i.e.} $\Phi_{\mathrm{um}}[\Hop_A] \neq \Hop_A$ \cite{Yi2017,Ding2018a}. Our first choice for the reference state of the sub-system $A$ is the fixed or invariant state of the above map, which we denote by $\hrho_{A,\mathrm{um}}$ with:
\begin{align}
	\Phi_{\mathrm{um}} [\hrho_{A,\mathrm{um}}] = \hrho_{A,\mathrm{um}} \label{eq:unmonRef}. 
\end{align}
From the definition of the map in Eq. \eqref{eq:unmonmap}, it is clear that the invariant state of the unmonitored cycle depends on both the unitary operator $\Uop$ and the measurement operator and unlike the thermal state of $B$ can also have coherences in the energy basis. 

Both for the purpose of evaluating the work and energy exchange statistics, as well as describing the second choice for the initial state of sub-system $A$ we now introduce projective measurements of energy of the total system $A+B$ before and after the unitary stroke. Let $\{\epsilon_{m_A},\epsilon_{m_B}\}$ and $\{\epsilon_{n_A},\epsilon_{n_B}\}$ denote the energy eigenvalues of the initial and final measurement respectively, and $\Piop_{m} = \Piop_{m_A} \otimes \Piop_{m_B}$ and $\Piop_{n} = \Piop_{n_A} \otimes \Piop_{n_B}$ the corresponding eigen-projectors.  The state of sub-system $A$ after the cycle with interspersed energy measurements is given by the following map:
\begin{align}
	\Phi_{\mathrm{m}} [\hrho_{A0}] =  \displaystyle \sum_{l_A,m_A} \Piop_{l_A} T_{\mathrm{cyc}}(l_A,m_A) \Tr_{A}[\Piop_{m_A} \hrho_{A0}] \label{eq:monmap}.
\end{align}
Here the transition matrix $T_{\mathrm{cyc}}(l_A,m_A)$ denotes the probability that system $A$ is in the state with energy $\epsilon_{l_A}$ after the cycle given it is initially in the state with energy $\epsilon_{m_A}$. This matrix can be written as:
\begin{align}
&T_{\mathrm{cyc}}(l_A,m_A) = \sum_{n_A,n_B,m_B} \Tr_{A} \left[\Piop_{l_A} \Phi_{\mathcal{M}}\left(\Piop_{n_A}\right) \right] \nonumber \\
&\Tr_{A+B} \left[ \Piop_{n_A} \otimes \Piop_{n_B} \Uop \Piop_{m_A} \otimes \Piop_{m_B} \Uop^{\dagger}\right]  p_{\beta}(m_B).
\end{align}
With this, we denote the invariant reference state for the monitored TSM heat engine cycle by $\hrho_{A,\mathrm{m}}$ and it satisfies the equation:
\begin{align}
	\Phi_{\mathrm{m}} [\hrho_{A,\mathrm{m}}] = \hrho_{A,\mathrm{m}} \label{eq:monRef}. 
\end{align}
In this approach as we subject the initial state to projective energy measurements, the invariant state has to be a diagonal state without coherences in the energy basis. As a result, we observe that the invariant state's population in the energy basis is nothing but the properly normalized eigenvector with eigenvalue $1$ of the reducible transition matrix $T_{\mathrm{cyc}}(l_A,m_A)$ \cite{Rezek2006,Dann2020,Ding2018a}. Comparing Eqs. \eqref{eq:unmonRef} and \eqref{eq:monRef}, it becomes clear that the invariant reference state in the monitored and unmonitored case can in general be different and consequently engine performance metrics such as the work output and its fluctuations can be rather different for them. We note that this was also pointed out in the context of a four stroke quantum Otto heat engine in a recent publication \cite{Son2021}. We will compare and contrast our findings with \cite{Son2021} in the conclusion.

Having described the two choices for the initial density matrix $\hrho_{A0}$ in Eqs. \eqref{eq:unmonRef} and \eqref{eq:monRef}, we now consider the work and energy exchange statistics given a particular initial state as in Eq. \eqref{eq:initstate}. As mentioned before, we adopt a TMA strategy as in previous works \cite{Campisi2015,Sacchi2021a,Sacchi2021b} and write the characteristic function corresponding to the joint probability distribution function $p(w,\Delta E_A)$ of the work input by the unitary, $w$,  and the energy change of sub-system $A$, $\Delta E_A$ as:
\begin{align}
	&\chi(\lambda,\mu)  = \int dw d \Delta E_A e^{i \lambda w+i\mu \Delta E_A} p(w,\Delta E_A) = \nonumber \\
	& \Tr [ \Uop^{\dagger} e^{i \mu \Hop_A} e^{i\lambda (\Hop_A+\Hop_B)}  \Uop e^{-i\lambda (\Hop_A+\Hop_B)} e^{ - i \mu \Hop_A}  \hrho_{0,d} ]  \label{eq:charfn_gen}
\end{align}
where $\hrho_d = \sum_j \Piop_j \hrho_0 \Piop_j$ is the projection of the initial density matrix to the energy eigenbasis. For the sake of completeness, we note that for the instance of TMA described earlier the stochastic variables $w = (\epsilon_{n_A}+\epsilon_{n_B})-(\epsilon_{m_A}+\epsilon_{m_B})$,  and $\Delta E_A = \epsilon_{n_A}-\epsilon_{m_A}$. The characteristic function $\chi(\lambda,\mu)$ then determines all the moments of the variables $w$ and $\Delta E_A$ as:
\begin{align}
	\avg{w^j \Delta E_A^k} = (-i)^{j+k} \left . \frac{\partial^{j+k} \chi(\lambda,\mu)}{\partial \lambda^j \partial \mu^k} \right \vert_{\mu=\lambda=0} \label{eq:moments} .
\end{align}
In particular, we note that the average work and energy change can be written as:
\begin{align}
	\avg{w} &= -i \left . \frac{d \chi(\mu,\lambda)}{d \lambda} \right \vert_{\mu=\lambda=0} \label{eq:averagew} \\
-\avg{E_{\mathcal{M}}} &= \avg{\Delta E_A} = -i \left . \frac{d \chi(\mu,\lambda)}{d \mu} \right \vert_{\mu=\lambda=0} \label{eq:averageDelE}.
\end{align}
Here, $E_{\mathcal{M}}$ denotes the energy input by the measurement and this along with the heat exchanged with the thermal bath $Q_c$, that satisfies $\avg{Q_c} = - \avg{\Delta E_B} = \avg{\Delta E_A-w}$, \footnote{Note that relations such as $Q_c = \Delta E_A - w$, $E_{\mathcal{M}} = -\Delta E_A$ at the level of the stochastic variables are not justified in general.} fuel the heat engine with the average work output given by $-\avg{w}$. In our analysis of particular realizations of TSM heat engines below, in addition to the average work, we are also interested in the reliability of the engine cycle defined as follows:
\begin{align}
	\mathrm{R} = \frac{-\avg{w}}{\sigma_w} \label{eq:Reliability},
\end{align}
with the standard deviation of work output given by $\sigma_w = \sqrt{\avg{w^2}-\avg{w}^2}$. As apparent from the definition of the characteristic function in Eq. \eqref{eq:charfn_gen}, we see that the off-diagonal elements or coherences of the initial density matrix $\hrho_{A0}$ (which can be non-zero for the unmonitored reference state) play no role in determining the work statistics \cite{Solinas2015,Talkner2016,Perarnau-Llobet2017}. As a measure of the average work output including the effect of coherences, we define 
the following variable which simply tracks the average energy change of the total system during to the unitary work stroke:
\begin{align}
	\avg{w}_c = \Tr_{A+B} \left [(\Hop_A+\Hop_B)(\Uop \hrho_0 \Uop^{\dagger} - \hrho_0) \right] \label{eq:coherentworkdef}. 
\end{align}

\section{TSM Heat Engine with Two Qubit Working Fuel}
\label{sec:sec3}
Having delineated the set-up of the TSM heat engine for a general bipartite system, from hereon we focus on a specific realization with the working fuel given by a two-qubit system with the hamiltonians:
\begin{align}
	\Hop_A = \frac{\omega_A}{2} \hs_{z}^{A},	\Hop_B = \frac{\omega_B}{2} \hs_{z}^{B} \label{eq:Hqubit},
\end{align}
with $\hs_{\tau}^{A}$ and  $\hs_{\tau}^{B}$ with $\tau = (x,y,z,\pm)$ denoting the usual Pauli and ladder operators for qubit $A$ and $B$ respectively (we set $\hbar = 1$ and $k_B = 1$ throughout). With this choice for the system we consider three successive choices for the unitary operator $\Uop$ and the measurement map $\Phi_{\mathcal{M}}$, beginning with a partial swap unitary and gaussian POVM. In each of these cases we will initially focus only on the behavior of the average work output and consider the reliability in the last sub-section.

\begin{figure*}
	\centering
	\subfloat[]{\includegraphics[width=0.33\linewidth]{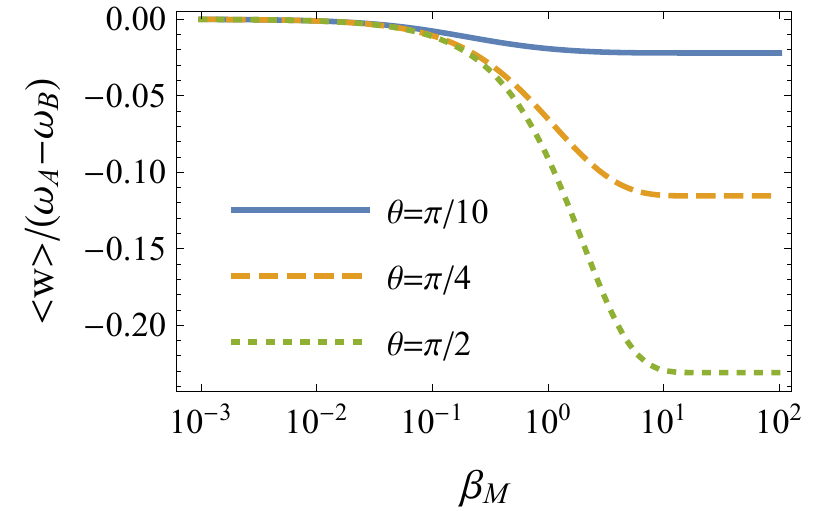}} \hfill
	\subfloat[]{\includegraphics[width=0.33\linewidth]{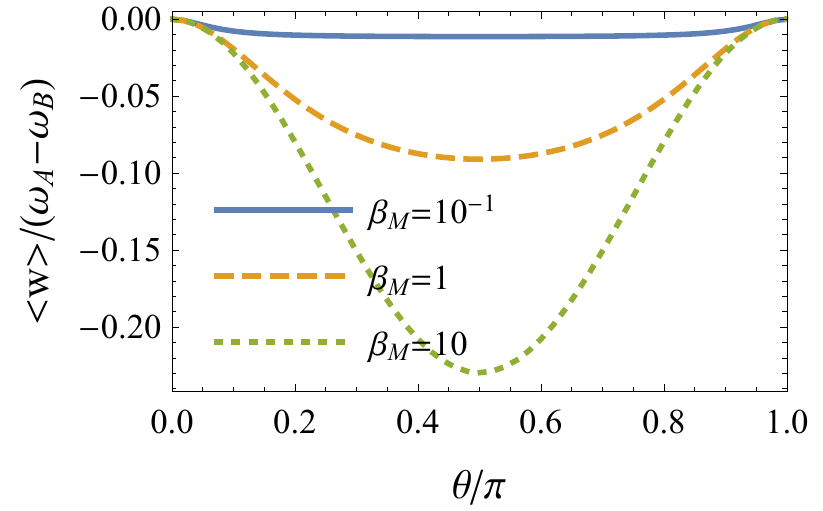}} \hfill
	\subfloat[]{\includegraphics[width= 0.33\linewidth]{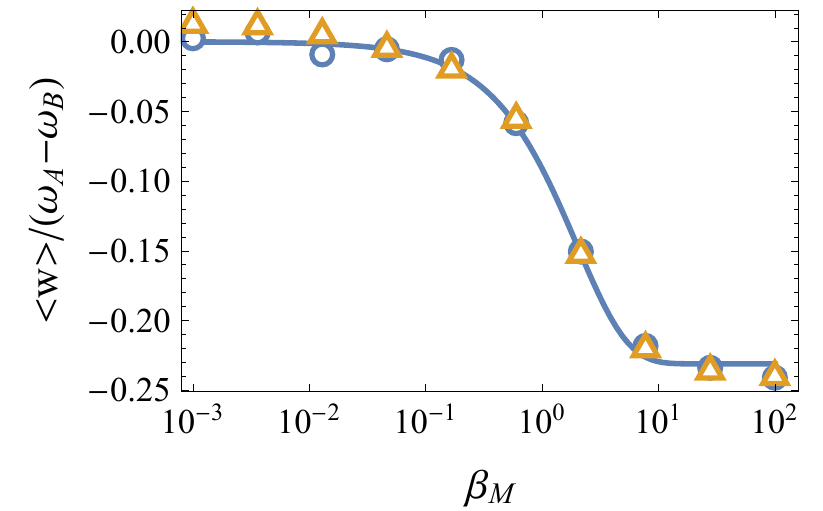}} 
	\caption{(Color Online) Average work output (gaussian POVM) as a function of measurement strength $\beta_M$ (a) and partial swap unitary parameter $\theta$ (b).  In (c) the average work output ($\theta = \pi/2$) is compared to a numerical simulation of multiple cycles of the TSM heat engine with (orange triangles) and without (blue circles) monitoring diagnostic measurements (see text for details).}
	\label{fig:figure2}
\end{figure*}
\subsection{Partial Swap Unitary and gaussian POVM Measurement}

For the first demonstration, we consider the following partial swap unitary \cite{Quan2007,Campisi2015,Sacchi2021a,Sacchi2021b}
\begin{align}
\Uop = \Uop_{\theta} = \exp \left ( i \theta [\hs_+^A \hs_-^B + \hs_-^A \hs_+^B] \right) \label{eq:Uqubit},
\end{align}
and the measurement map is given in terms of the following gaussian POVM with continuous outcomes $-\infty < q < \infty$:
\begin{align}
	\Phi_{\mathcal{M}}[\hrho] &= \int_{-\infty}^{\infty} dq \hM_q(q_0,\sigma) \hrho \hM_q(q_0,\sigma) \label{eq:gaussmeasop}\\
	\hM_{q}(q_0,\sigma) &= \frac{1}{(2\pi\sigma^2)^{1/4}} \left( e^{-\frac{(q-q_0)^2}{4\sigma^2}} \Piop_{x_+,A} \right . \nonumber \\
	& \left .  +e^{-\frac{(q+q_0)^2}{4\sigma^2}}\Piop_{x_-,A}\right)  \label{eq:gaussmeasopqubit},
\end{align}
where $\Piop_{x_\pm,A} = \ket{x_\pm}\bra{x_\pm} $ are projectors corresponding to the eigenstates of $\hs_x^A$ \emph{i.e.} $\hs_x^A \ket{x_\pm} = \pm \ket{x_\pm}$. This POVM clearly signifies a weak measurement of the operator $\hs_x^A$. Using the forms of unitary operator and measurement from Eqs. \eqref{eq:Uqubit} - \eqref{eq:gaussmeasopqubit}, we solve Eqs. \eqref{eq:monRef} and \eqref{eq:unmonRef} for the invariant reference states in the monitored and unmonitored cycles. In this case, we find that the solution for the monitored and unmonitored case are identical \emph{i.e.} 
\begin{align}
		\hrho_{A,\mathrm{m}}=\hrho_{A,\mathrm{um}} = \begin{pmatrix}
				\popa & 0\\
				0 & 1-\popa
			\end{pmatrix} \label{eq:gaussMrho},
\end{align}
with the probability $\popa$ satisfying:
\begin{align}
\popa = \popb -2 \frac{(1/2-\popb)(e^{\beta_M/2}-1)}{1-2e^{\beta_M/2}+\cos2 \theta} \label{eq:gausspop}.
\end{align}
Here $\popb = 1/(1+e^{\beta \omega_B})$ denotes the population of the excited state in the initial thermal state of qubit $B$ and $\beta_M = q_0^2/\sigma^2$ is a dimensionless variable characterizing the strength of the gaussian POVM. Since $\popb < 1/2$, we immediately notice that  Eq. \eqref{eq:gausspop} guarantees $\popa \geq \popb$ - establishing the measurement `bath' as an analogue to the hot bath in the usual two-stroke heat engines considered in \cite{Quan2007,Campisi2015,Sacchi2021a,Sacchi2021b}. This could also have been anticipated from the fact that the measurement inputs energy in general \cite{Yi2017}. From Eq. \eqref{eq:gausspop}, we find that  when the measurement strength is really weak with $\beta_M \rightarrow 0$, we have that $\popa \rightarrow \popb$ and in the opposite limit of a strong measurement with $\beta_M \rightarrow \infty$, we have $\popa \rightarrow 1/2$. Thus the parameter $\beta_M$ controls the effective temperature of the ``hot"-measurement bath and by changing $\beta_M$, we can tune the effective temperature of $A$ from the temperature of the cold bath $\beta$ to infinite temperature. With this, for the partial swap unitary we can naturally write down the average work, for both monitored and unmonitored cases, as \cite{Quan2007,Campisi2015,Sacchi2021a,Sacchi2021b}:
 \begin{align}
 	\avg{w} &= -(\popa-\popb) (\omega_A-\omega_B)\sin^2 \theta \label{eq:wavgqubitGaussPOVM}.
 \end{align}
\begin{figure*}
	\centering
	\subfloat[]{\includegraphics[width=0.33\linewidth]{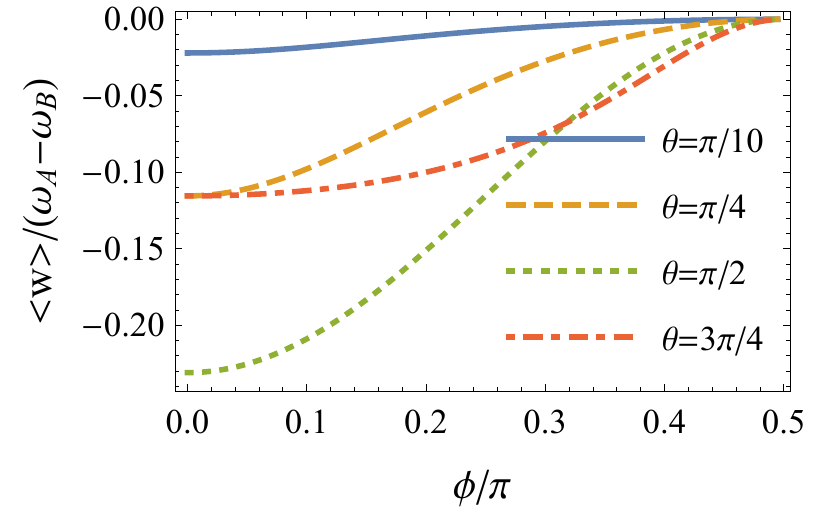}} \hfill
	\subfloat[]{\includegraphics[width=0.33\linewidth]{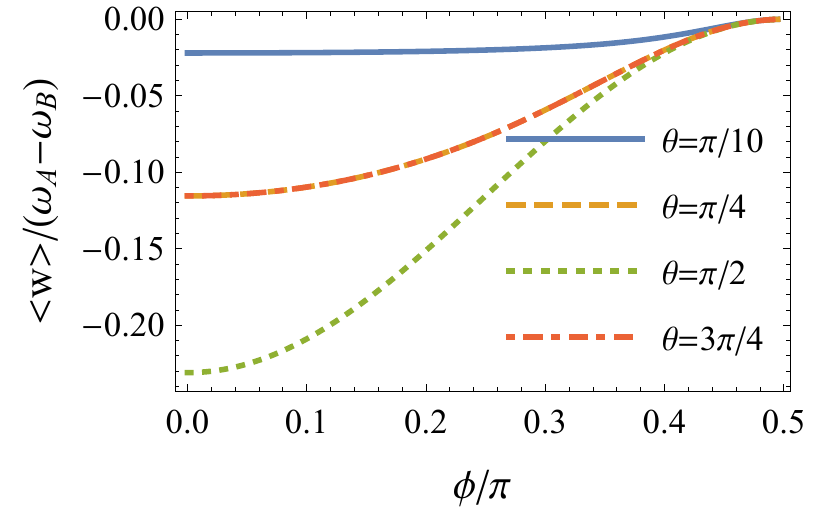}} \hfill
	\subfloat[]{\includegraphics[width= 0.33\linewidth]{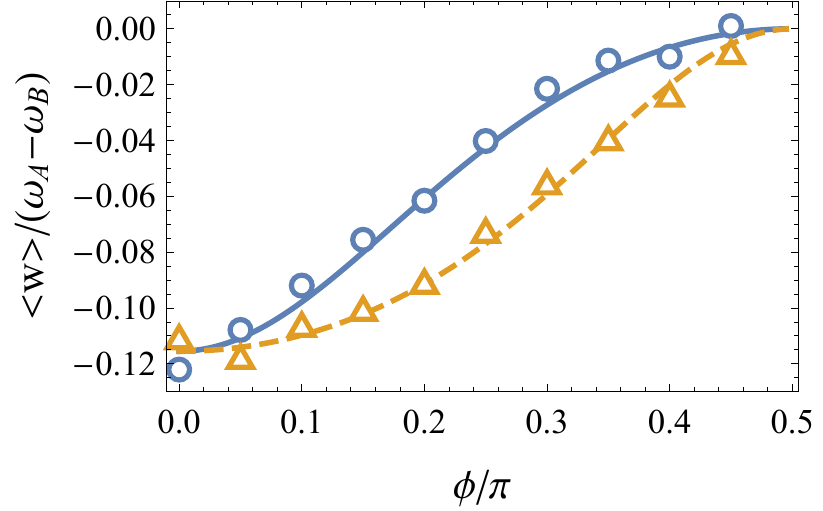}} 
	\caption{(Color Online) Average work output (with projective measurement) as a function of measurement direction $\phi$ for unmonitored TSM heat engine cycle (a) and monitored TSM heat engine cycle (b). In (c) the average work output (lines) iis compared to a numerical simulation of multiple cycles of the TSM heat engine with (orange triangles and dashed lines) and without (blue circles and solid lines) monitoring diagnostic measurements ($\theta = \pi/4$).}
	\label{fig:figure3}
\end{figure*}
Clearly, as long as $\omega_A>\omega_B$, we always get a positive average work output \emph{i.e.} $\avg{w}<0$. As depicted in Fig. \eqref{fig:figure2} (a), the average work output is monotonically increasing as a function of the measurement strength $\beta_M$ and the highest value of work output is achieved with the perfect swap unitary with $\theta = \pi/2$ as evident from Fig. \eqref{fig:figure2} (b). We have so far examined the average work output with the sub-system $A$ prepared in a given initial state. Since the initial states of interest in Eqs. \eqref{eq:unmonRef} and \eqref{eq:monRef} are invariant reference states of the monitored and unmonitored cycle, there is yet another way to realise such initial states in practice. We can simply begin with the sub-system $A$ in an arbitrary state and run the TSM heat engine through multiple cycles. After a significant number of cycles the system is expected to settle down to the reference state. In order to confirm this behavior we perform numerical simulation of the engine operation by running the TSM heat-engine over $20$ cycles. For the unmonitored case, to calculate the average work from the final invariant state reached by the engine, we simulate the diagnostic work TMA measurement $20,000$ times. In the monitored case, we simulate $20000$ trajectories of $20$ cycles with diagnostic measurements during the cycles to calculate the average work. The process we described for the numerical validation is kept the same throughout and in all the results presented in Figs. \eqref{fig:figure2}-\eqref{fig:figure7}, the cold bath inverse temperature is taken as $\beta = 1/\omega_B$. As expected, for the gaussian POVM measurement we can see from Fig. \eqref{fig:figure2} (c) that the monitored and unmonitored cases give exactly the same value of $\avg{w}$ and agree well with the analytical result.

\subsection{Partial Swap Unitary and Projective Measurement}

The gaussian POVM measurement operators considered in the previous sub-section gave the same diagonal invariant reference state for both the monitored and unmonitored cases. We now consider an alternate choice for the measurement that will lead to a rather different result. To that end, consider a projective measurement of the qubit spin-projection along the direction $\nn = \cos(\phi) \mathbf{e}_x + \sin(\phi) \mathbf{e}_z$ (with $\mathbf{e}_{x(z)}$ unit vector along $x(z)$ direction) giving the measurement map:
\begin{align}
	\Phi_{\mathcal{M}}[\hrho] &= \sum_{k=\pm}  \Piop_{\nn_k,A}\hrho \Piop_{\nn_k,A} \label{eq:projmeasmap}\\
     \Piop_{\nn_k,A} & = \ket{\nn_k}\bra{\nn_k}  \label{eq:projmeasoper},
\end{align}
with $\hs_{\nn}^A\ket{\nn_k}= k \ket{\nn_k}$ and $k=\pm$, and $\hs_{\nn}^A = \hs_x^A \cos(\phi) + \hs_z^A \sin(\phi)$. Keeping the same partial-swap unitary (see Eq. \eqref{eq:Uqubit}) as before, we now solve Eqs. \eqref{eq:unmonRef} and \eqref{eq:monRef} for the reference states. In the unmonitored case, we find that the reference state $\hrho_{A,\mathrm{um}}$ is not diagonal but can have real valued coherences in the energy basis. The population of the excited state of $\hrho_{A,\mathrm{um}}$, $\popaum$, and the coherence, $c_R$, take the form:
\begin{align}
	\popaum &=  \popb + \frac{(1-2\popb) \cos^2(\phi)}{2+2\cos(\theta)\sin^2(\phi)}\label{eq:popunmonU1Proj},\\
	c_R & = \frac{(-1+2\popb) \cos^2(\theta/2)\sin(2\phi)}{2+2\cos(\theta)\sin^2(\phi)} \label{eq:cohunmonU1Proj}.
\end{align}
On the other hand, in the monitored case the reference state $\hrho_{A,\mathrm{m}}$ continues to be diagonal in the energy basis with the population of the excited state $\popam$ given by:
\begin{align}
	\popam = \popb + \frac{2(1-2\popb)\cos^2(\phi)}{3+\cos(2\phi)-2\cos(2\theta) \sin^2(\phi)}\label{eq:popmonU1Proj}.
\end{align}	
\begin{figure*}
	\centering
	\subfloat[]{\includegraphics[width=0.33\linewidth]{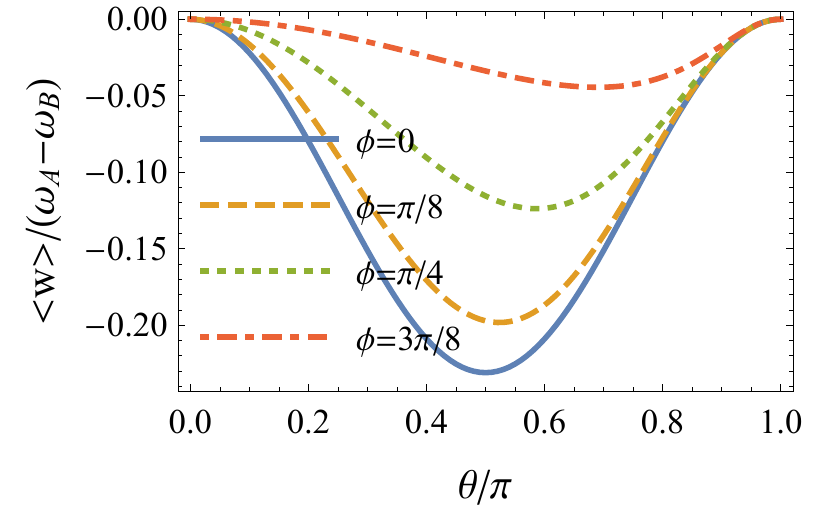}} \hfill
	\subfloat[]{\includegraphics[width=0.33\linewidth]{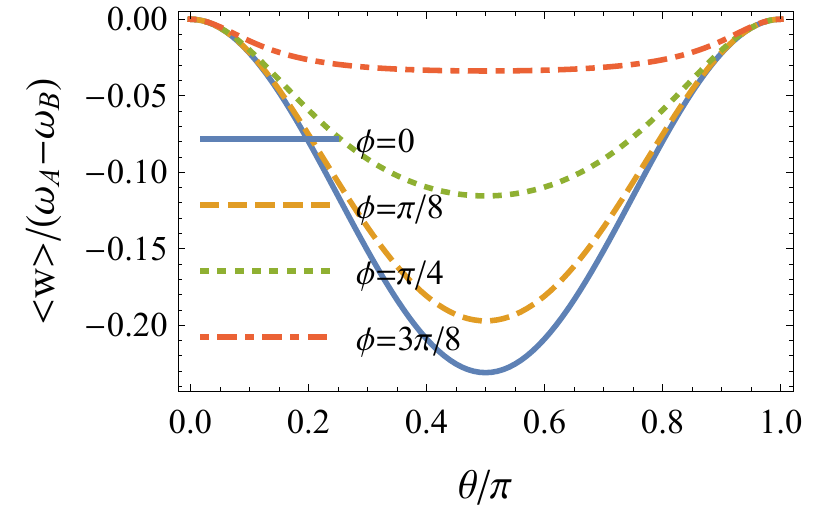}} \hfill
	\subfloat[]{\includegraphics[width= 0.33\linewidth]{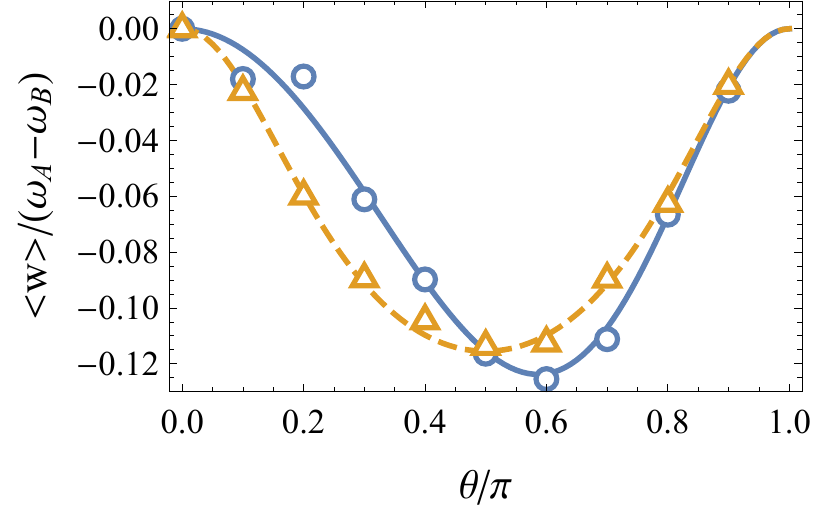}} 
	\caption{(Color Online) Average work output (with projective measurement) as a function of partial swap unitary parameter $\theta$ for unmonitored TSM heat engine cycle (a) and monitored TSM heat engine cycle (b). In (c) the average work output (lines) is compared to a numerical simulation of multiple cycles of the TSM heat engine with (orange triangles and dashed lines) and without (blue circles and solid lines) monitoring diagnostic measurements ($\phi=\pi/4$).}
	\label{fig:figure4}
\end{figure*}
As before by inspection we can immediately see that in both the unmonitored and monitored cases we have $\popaum\geq \popb$ and $\popam \geq \popb$ making the measurement act as an effective hot bath. A central result of our paper is that the populations in general satisfy $\popaum \neq \popam$ leading to distinct TMA work statistics for the monitored and unmonitored case. We note that this has nothing to do with the non-zero coherence $c_R$ since coherences do not play a role in determining work statistics in the TMA approach. Since we have used the swap unitary as in the previous sub-section the average work output takes the same form as in Eq. \eqref{eq:wavgqubitGaussPOVM} with the $\popa$ replaced for the unmonitored and monitored population discussed above giving
\begin{align}
	\avg{w} &= -(\popaum-\popb) (\omega_A-\omega_B)\sin^2 \theta \label{eq:wavgqubitProjUnMon},
\end{align}
for the unmonitored cycle and
\begin{align}
	\avg{w} &= -(\popam-\popb) (\omega_A-\omega_B)\sin^2 \theta \label{eq:wavgqubitProjMon},
\end{align}
for the monitored cycle. Thus, as before, $\omega_A>\omega_B$ ensures that we have engine operation with positive average work output. In Figs. \eqref{fig:figure3} and \eqref{fig:figure4} we have explored the behavior of the average work output of monitored and unmonitored cycles. A key point that is apparent from Eqs. \eqref{eq:popunmonU1Proj} and \eqref{eq:popmonU1Proj} and reflected in Figs. \eqref{fig:figure3} (a,b) is that the angle $\phi$ can be used to control the extent to which the projectors $ \Piop_{\nn_k,A}$ commute with the hamiltonian $\Hop_A$ and consequently the average energy input from the measurement. For a given unitary stroke ($\theta$ - fixed), invariably the maximum work output is for the $\phi = 0$ case which leads to $\popaum=\popam = 1/2$ (infinite effective temperature measurement bath) and the work output goes to zero with $\popaum=\popam = \popb$ for $\phi = \pi/2$.  From Fig. \eqref{fig:figure4} (a,b) depicting the average work output as a function of $\theta$ it becomes clear that for the unmonitored case, for a given $\phi$, the maximum work output is not always at $\theta = \pi/2$ unlike the monitored case. In fact this is also apparent from Fig. \eqref{fig:figure3} (a). This can be understood from Eq. \eqref{eq:popunmonU1Proj} where $\popaum$ does not have reflection symmetry about $\theta = \pi/2$. Consequently, the relative magnitudes of the average work output in unmonitored or monitored case varies with the unitary ($\theta$) and the measurement operator ($\phi$) chosen. By examining the difference between the average work in monitored and unmonitored case, we find that the work output is larger in the monitored case for $\theta<\pi/2$ and vice-versa for $\theta>\pi/2$. This is illustrated in Figs. \eqref{fig:figure3} (c) and \eqref{fig:figure4} (c) where we compare the monitored and unmonitored work output for some exemplary values of $\theta$ and $\phi$. In addition, we also note that many cycle, multiple trajectory simulation of the TSM engine cycle (circles and triangles) agree very well with the plotted analytical results.

Since the unmonitored reference density matrix has non-zero coherences (see Eq. \eqref{eq:cohunmonU1Proj}), one may suspect that in this case the coherent average work defined in Eq. \eqref{eq:coherentworkdef} may lead to a different prediction compared to the TMA work given by Eq. \eqref{eq:wavgqubitProjUnMon}. We can easily check that this is not the case and we have $\avg{w}_c = \avg{w}$. This ultimately comes from the nature of the partial swap unitary Eq. \eqref{eq:Uqubit} that does not couple populations and coherences. In the next sub-section, we address this issue with a different choice for the unitary.

\subsection{Augmented Partial Swap Unitary and Projective Measurement}
\begin{figure*}
	\centering
	\subfloat[]{\includegraphics[width=0.33\linewidth]{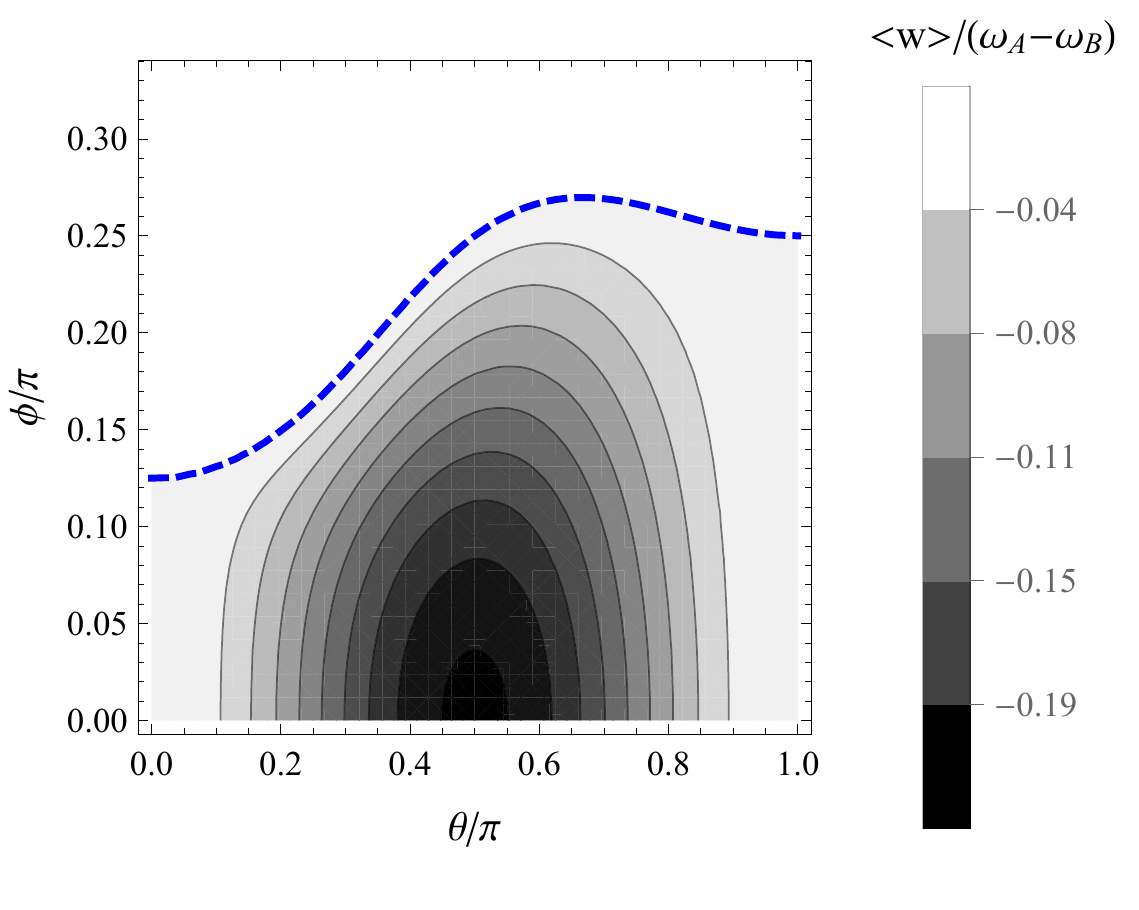}} \hfill
	\subfloat[]{\includegraphics[width=0.33\linewidth]{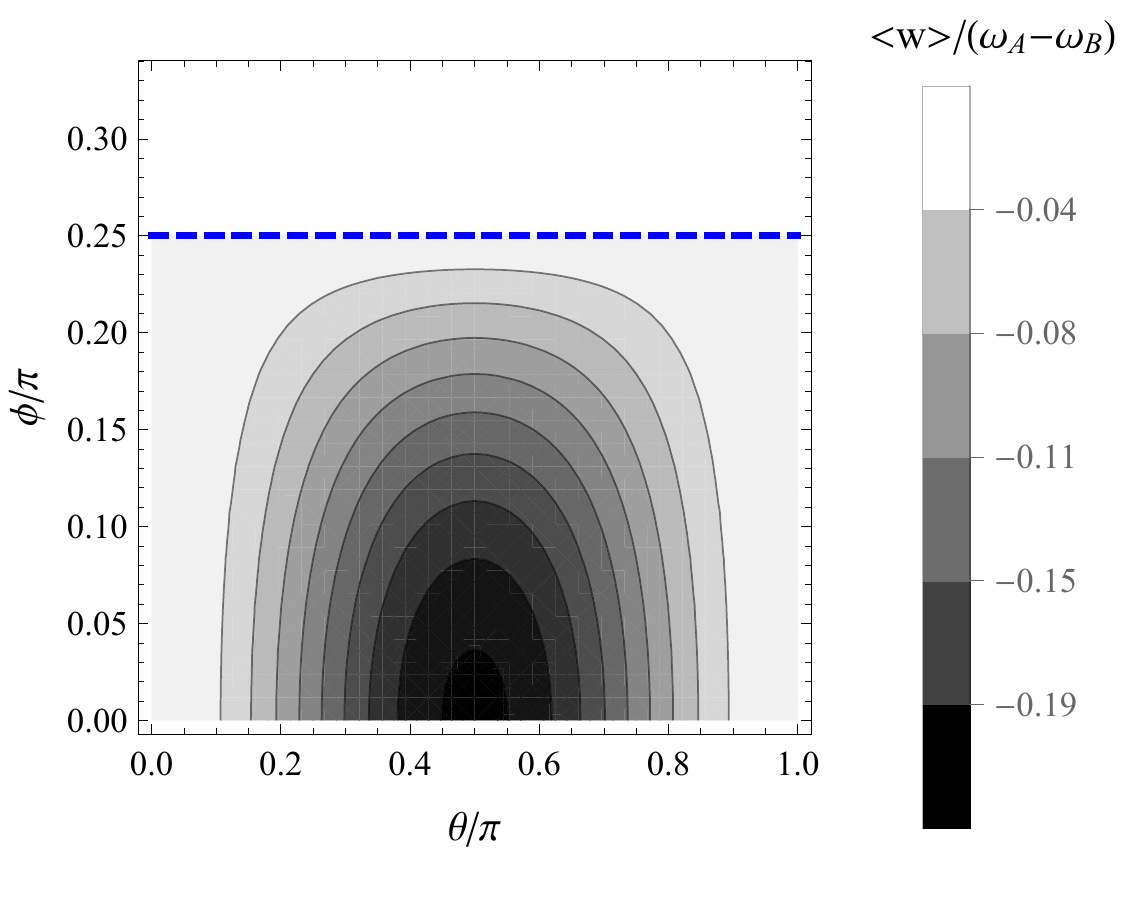}} \hfill
	\subfloat[]{\includegraphics[width= 0.33\linewidth]{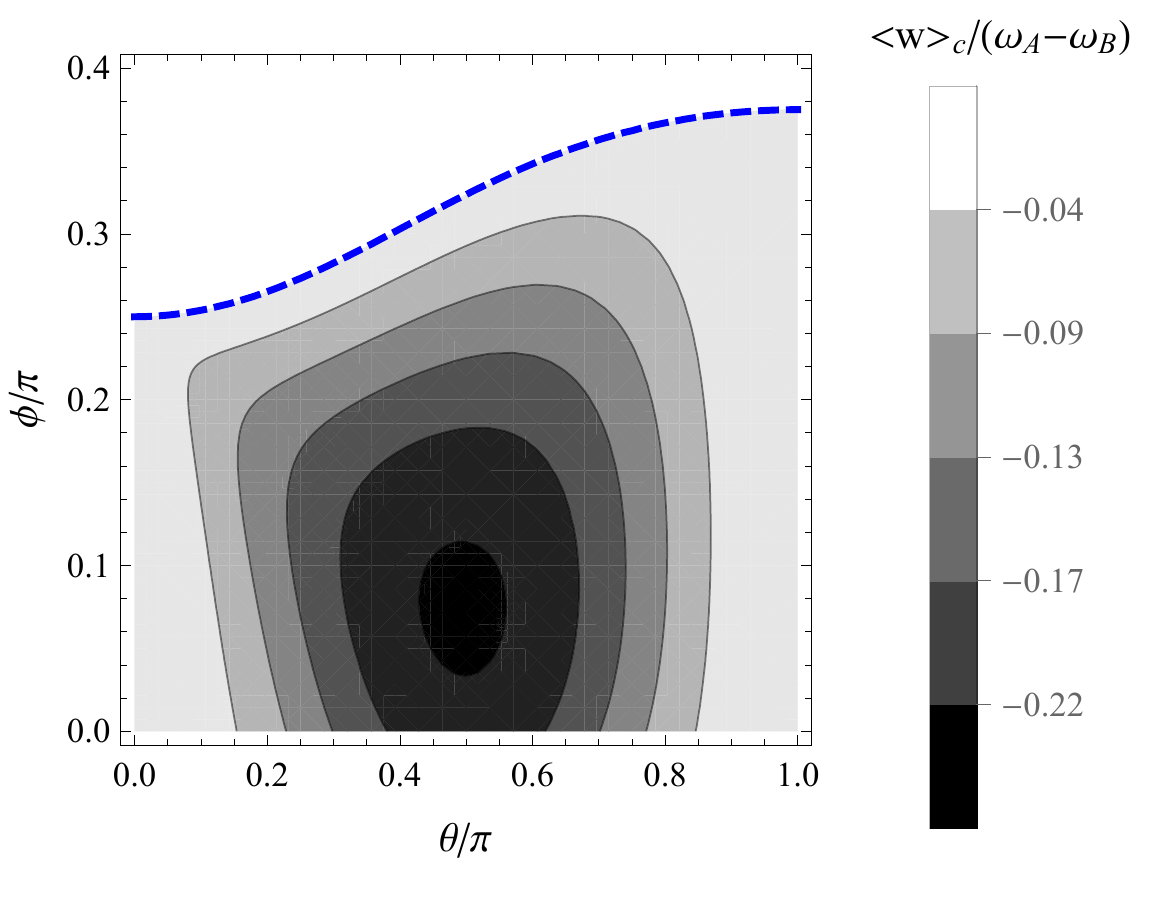}} 
	\caption{(Color Online) Average work output (with projective measurement) as a function of measurement direction $\phi$ and parameter $\theta$ characterizing the partial swap unitary augmented with a Hadamard operator for (a) unmonitored TSM heat engine cycle, (b) monitored TSM heat engine cycle, and (c) coherent average work without TMA measurements. Dashed blue lines indicate the boundary above which the work output becomes non-positive. $\omega_A=2 \omega_B$ here.}
	\label{fig:figure5}
\end{figure*}
In order to allow the possibility of the coherences in the unmonitored reference state to play a role in determining  the coherent average work $\avg{w}_c$, we consider the following unitary operator
\begin{align}
	\Uop = \Uop_{\theta,H} =  \Uop_{\theta} \Uop_H  \label{eq:Uqubitcoh},
\end{align}
with $\hat{H}$ representing the Hadamard operator:
\begin{align*}
	\hat{H} = \ket{x_+}\bra{z_+} \, +\, \ket{x_-}\bra{z_-},
\end{align*}
with $\hs_z \ket{z,\pm} = \pm \ket{z,\pm}$. Thus, $ \Uop_{\theta,H}$ is essentially the partial swap operator augmented with the Hadamard operator.  We choose $\Uop_{\theta,H}$ for the unitary work stroke and take the projective measurement operator in Eq. \eqref{eq:projmeasoper} to construct the cycle and solve equations \eqref{eq:unmonRef} and \eqref{eq:monRef} to determine the reference states. As in the previous section, we find that the density matrix in the unmonitored case has non-zero coherence $c_R$. The population of the excited state and coherence for the unmonitored cycle's reference state $\hrho_{A,\mathrm{um}}$ are given by:
\begin{align}
	\popaum &= \frac{f(p_\beta,\theta,\phi)}{-4+2 \cos(\theta)[1+\cos(\theta)]\sin(2\phi)} \label{eq:popunmonU2Proj}\\
	c_R &= \frac{(1-2\popb)\sin^2(\theta)\sin(2\phi)}{-4+2 \cos(\theta)[1+\cos(\theta)]\sin(2\phi)} \label{eq:cohunmonU2Proj},
\end{align}
with the somewhat cumbersome function $f(p_\beta,\theta,\phi) = [-3-2\popb+(-1+2\popb)(\cos 2\theta +2\cos 2\phi\,\sin^2\theta)+2(1+\cos 2 \theta)\cos \theta \, \sin 2\phi]/2$, and the population for the monitored reference state  $\hrho_{A,\mathrm{um}}$ is:
\begin{align}
	\popam &= [3+2\popb+(1-2\popb)(\cos 2\theta +2\cos 2\phi\,\sin^2\theta)]/8. \label{eq:popmonU2Proj}
\end{align} 
Clearly we see once again that the monitored and unmonitored cycles lead to very different reference states. Since we are using an augmented swap unitary operator here, we have to reevaluate the work statistics according to TMA. Upon this we find that the average work (for both monitored and unmonitored cases) \cite{Quan2007,Campisi2015,Sacchi2021a,Sacchi2021b} takes the following form:
\begin{align}
	\avg{w} &= -(\popa-\popb) (\omega_A-\omega_B)\sin^2 \theta \nonumber\\ 
	&+ (1/2-\popa)(\omega_A\cos^2\theta + \omega_B \sin^2 \theta)\label{eq:wavgTMAU2},
\end{align}
with $\popa$ chosen as $\popaum$ or $\popam$. Unlike the previous sub-sections, here $\omega_A>\omega_B$ alone does not guarantee positive average work output. Before we come to the condition for positive work output, in line with the key motivation for introducing the unitary operator $\Uop_{\theta,H}$, we find that the coherent average work (see Eq. \eqref{eq:coherentworkdef}) is given by:
\begin{align}
	\avg{w}_c &= -(\popaum-\popb) (\omega_A-\omega_B)\sin^2 \theta \nonumber\\ 
&+ (1/2-\popaum+c_R)(\omega_A\cos^2\theta + \omega_B \sin^2 \theta)\label{eq:wavgcohU2}.
\end{align}
As anticipated the coherent average work output has explicit dependence on the coherence $c_R$ in this case. From Eqs. \eqref{eq:popunmonU2Proj}- \eqref{eq:wavgcohU2}, we determine the conditions for the average work output to be positive for the unmonitored (TMA) cycle as
\begin{align}
	\frac{\omega_B}{\omega_A} \leq \frac{2\cos^2 \phi-\cos \theta (1+\cos \theta)\sin 2\phi}{2-\cos \theta (1-\cos \theta)\sin2\phi} \label{eq:PWCunmon},
\end{align}
monitored cycle as
\begin{align}
	\frac{\omega_B}{\omega_A} \leq  \cos^2 \phi \label{eq:PWCmon},
\end{align}
and for the coherent average work as
\begin{align}
		\frac{\omega_B}{\omega_A} \leq \frac{\cot \phi - \cos \theta}{\tan \phi + \cot \phi - (1+\cos \theta)} \label{eq:PWCcoh}.
\end{align}
One can view the above conditions as a stricter bound on $\omega_B/\omega_A$ than the condition $\omega_B < \omega_A$ which was enough to ensure positive work output in the cycle with the swap unitary. Alternatively, as we depict in Figs. \eqref{fig:figure5} (a,b,c), at a given value of $\omega_B/\omega_A$ the conditions above mark out the regions of $\theta,\phi$ for which the average work output is positive. In addition, comparing Fig. \eqref{fig:figure5} (b) with (a,c) we observe that in the unmonitored and coherent cases, the value of $\theta$ at which the work output is maximum is not necessarily $\theta = \pi/2$. This is in line with what we found in the swap unitary case before. 
\begin{figure}
	\includegraphics[width=0.8\linewidth]{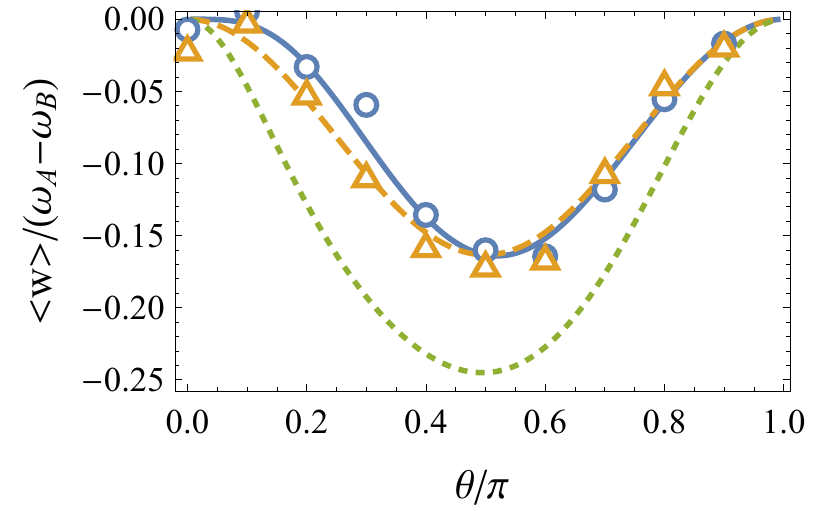}
	\caption{(Color online). Average work output (with projective measurement) as a function of parameter $\theta$ characterizing the partial swap unitary augmented with a Hadamard operator for  measurement direction $\phi = \pi/8$.  Solid blue line is for the unmonitored cycle, dashed orange represents the monitored cycle, and the dotted green line is the coherent average work. Analytical results are compared to a numerical simulation of multiple cycles of the TSM heat engine with (orange triangles) and without (blue circles) monitoring diagnostic measurements. $\omega_A=2 \omega_B$ here.} 
	\label{fig:figure6}
\end{figure}
Finally, we can now compare the different average work output measures. By examining the expressions in Eqs. \eqref{eq:wavgTMAU2} and \eqref{eq:wavgcohU2}, we can show that the TMA average work for the unmonitored cycle (denoted by $\avg{w}_{\mathrm{um}}$ ) is always greater than the coherent work \emph{i.e.} $\avg{w}_{\mathrm{um}}-\avg{w}_c \geq 0$. Thus, in the regime with positive work output with both $\avg{w}_c < \avg{w}_{\mathrm{um}} \leq 0$. Therefore, the coherent work average produces a larger average output than the case with TMA as depicted in Fig. \eqref{fig:figure6}.  As before the relative size of the average work output in the monitored and unmonitored case depends on the choice of $\theta$ and $\phi$. Moreover depending on the ratio of $\omega_A/\omega_B$ the coherent average work can be greater or lesser than the monitored case. For the choice of $\omega_A/\omega_B = 2$ made in Figs. \eqref{fig:figure5} and \eqref{fig:figure6}, we have that the coherent average work output is the largest.
\begin{figure*}
	\centering
	\subfloat[]{\includegraphics[width=0.33\linewidth]{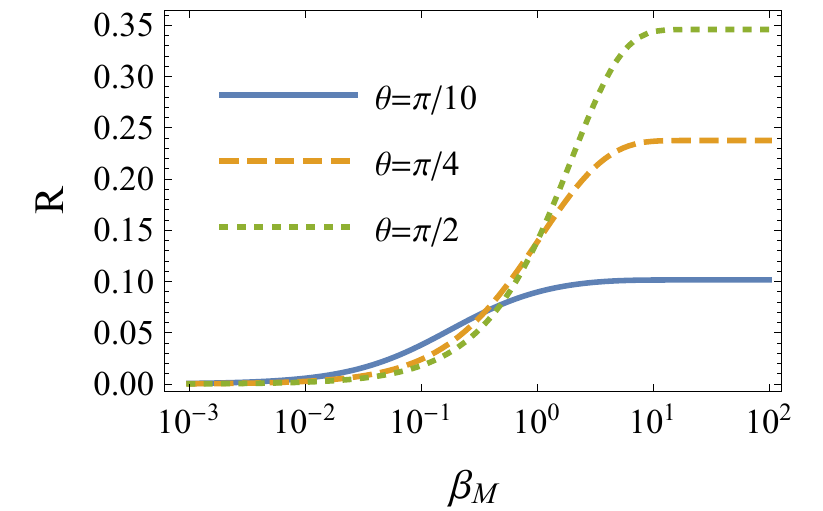}} \hfill
	\subfloat[]{\includegraphics[width=0.33\linewidth]{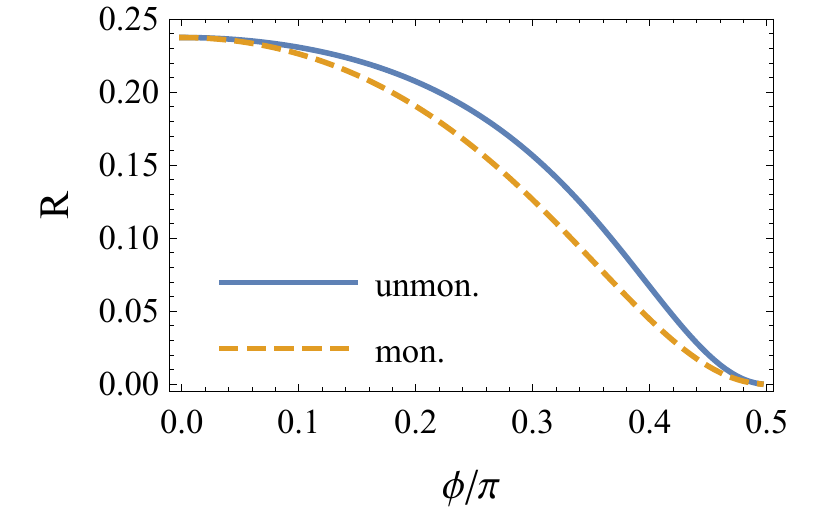}} \hfill
	\subfloat[]{\includegraphics[width= 0.33\linewidth]{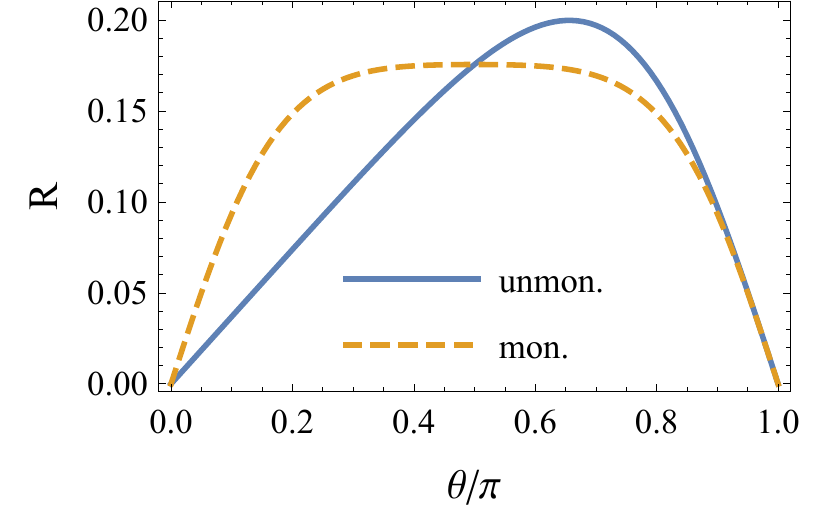}} 
	\caption{(Color Online) Reliability of the TSM Heat Engine cycle for gaussian POVM measurement (a) as a function of measurement strength and projective measurement with partial swap unitary as a function of measurement direction $\phi$ ($\theta = 3 \pi/4$) (b) and unitary parameter $\theta$ ($\phi = \pi/3$)(c). In (b,c) solid blue lines are for the unmonitored cycle and orange dashed line are for the monitored case.}
	\label{fig:figure7}
\end{figure*}
\subsection{Reliability}
Having extensively discussed the behavior of the average work output, we briefly consider the fluctuations in the work output. This aspect has become rather important both from the practical point of view of designing thermal machines with minimum noise in the output as well as in the context of understanding the fundamental limits of fluctuations in thermal machines as encoded by the recently discovered thermodynamic uncertainty relations (TUR) \cite{Barato2015,Timpanaro2019,Sacchi2021a,Sacchi2021b}. To this end, focusing on the swap unitary Eq. \eqref{eq:Uqubit}, we can use the characteristic function in Eq. \eqref{eq:charfn_gen} to write down the variance in work as:
\begin{align}
	\sigma_w^2 &= (\omega_A-\omega_B)^2 \sin^2\theta \left[\popa+\popb-2\popa\popb \right . \nonumber \\
	& \left . -  (\popa-\popb)^2 \sin^2 \theta \right]. \label{eq:TMAvariance}
\end{align}
Combining the above with Eq. \eqref{eq:wavgqubitGaussPOVM} we can easily obtain the reliability $R$ defined in Eq. \eqref{eq:Reliability}. The exact expressions for $R$ in the different scenarios discussed before can be obtained by substituting for the populations in the reference state obtained in Eqs. \eqref{eq:gausspop}, \eqref{eq:popunmonU1Proj}, and \eqref{eq:popmonU1Proj}. The resulting expressions for $R$ are cumbersome and we do not present them here. Instead, we have plotted the results for the reliability for some exemplary cases in Fig. \eqref{fig:figure7}. In the case of gaussian POVM, Fig. \eqref{fig:figure7} (a) illustrates that the reliability in general increases with the strength of the measurement parameterized by $\beta_M$. Moreover for strong measurements, the largest reliability occurs at the value $\theta = \pi/2$ where the work output is maximized. For weaker measurement $\beta_M \ll 1$, the reliability is high for small $\theta$ and decreases as $\theta$ is increased towards $\pi/2$. Thus for TSM engine cycles with weaker measurement, the most reliable choice for the unitary is not the one producing the maximum average output. Fig. \eqref{fig:figure7} (b) and (c) we compare the reliability of unmonitored and monitored cycles for the projective measurement (see Eq. \eqref{eq:projmeasoper}) case. The behavior of reliability essentially tracks the behavior of average work with regions of large average work output leading to higher reliability. Along the same lines as evident from Fig. \eqref{fig:figure7} (c), at a given value of $\phi$, the unmonitored case has larger reliability when $\theta>\pi/2$. We note that the behavior of reliability for the augmented swap unitary case discussed in previous sub-section is qualitatively similar to the swap unitary case. We also do not go into an extensive discussion of the TURs \cite{Timpanaro2019,Sacchi2021b}, except to note that they will always be satisfied for the two-qubit realizations of the TSM heat engine we have considered here. This is because we can always introduce an effective temperature corresponding to the reference state of the qubit system following which the TUR discussion will essentially mirror those presented in earlier publications \cite{Timpanaro2019,Sacchi2021a,Sacchi2021b}.

\section{Conclusion}
\label{sec:sec4}
In this paper, we have introduced and analyzed the working of a two-stroke measurement heat engine. The central idea is to replace one of the thermal reservoirs in previously studied models of two-stroke heat engines with bipartite working fuels \cite{Campisi2015,Sacchi2021a,Sacchi2021b} with a non-selective quantum measurement of one of the sub-systems (sub-system $A$). A distinct feature of the TSM heat engine is that the invariant or steady state of the sub-system subject to the measurement ``bath''  has to be self-consistently determined by solving equations Eqs. \eqref{eq:unmonRef} and \eqref{eq:monRef}. Consequently this state depends not just on the nature of the non-selective measurement map but also on the unitary implementing the work stroke and whether or not one performs diagnostic measurements during the cycle. This is the main feature that we have explored here. This dependence of the invariant reference state on the unitary work stroke is not new and has been explored in the context of quantum Otto cycles with finite thermalization time \cite{Rezek2006,Dann2020,Ding2018a}. Here, we have added to this line of research by understanding clearly the dependence of the reference state on whether the cycle is monitored or unmonitored. This is complementary to recent work \cite{Son2021} where it has been shown that in the context of a quantum Otto engine with multiple cycles the way one monitors the engine over many cycles strongly affects the total work output. In addition to this central aspect we have also identified that the unmonitored invariant state can have coherences and consequently lead to a different value of the average energy change in the unitary stroke (which we have called as coherent average work) than the average work from the TMA approach \cite{Solinas2015,Talkner2016,Perarnau-Llobet2017}.

Two-stroke thermal machines not only provide exemplary, simple, and clear realizations of a cyclic machine with bipartite quantum systems as working fuel, but as proposed in \cite{Campisi2015} they can also be implemented in current state-of-art superconducting qubit systems. One of the advantages in attempting the implementation of the TSM heat engine is that one needs to only implement a single thermal reservoir. In fact the gradients of (effective) temperature and consequently work output achievable with the TSM heat engine also make them more attractive than conventional two-stroke heat engines where one may face limitations in designing an experiment with a quantum system interacting with two baths with large difference in temperature. In the proposal in \cite{Campisi2015}, the qubits are given by Copper-pair boxes or flux qubits, the unitary work stroke is implemented by introducing an ancillary qubit, and two resistors maintained at given fixed temperatures play the role of thermal reservoirs \cite{Pekola2015,Pekola2019}. Given this proposal, the only change required to implement the TSM heat engine with qubits is to replace one of the resistor thermal baths with a POVM measurement that can be for instance implemented by reading out the qubit by coupling it to a resonator or waveguide. Indeed this extra step is well within current technological achievements as evident from the rich body of work studying quantum measurement with superconducting qubits in general \cite{Murch2013,Weber2014,Murch2016} and as test-bed for quantum thermodynamics \cite{Naghiloo2020}. 

In addition to a more detailed proposal for the implementation of the TSM heat engine, several interesting questions arise from our study. As we have remarked earlier, for the qubit system within the TMA approach since only the populations play a role we can map the TSM heat engine to a regular two-stroke heat engine by assigning an effective temperature to the measurement bath. An interesting direction would be to consider higher dimensional quantum systems with inhomogeneous spectrum and simple infinite dimensional quantum systems for the WF such as the harmonic oscillator \cite{Sacchi2021a} . We have preliminary results that suggest that even for a simple qutrit system the TSM heat engine can have reference states that cannot be given an unique effective temperature. While we have illustrated our idea via some exemplary choices for the unitary work strokes, the question of optimizing to find the unitary with the maximum work output for a given realisation of the measurement as done for the usual two-stroke heat engine \cite{Campisi2015} is interesting to pursue as well. Another line of inquiry of interest is to examine how a more detailed model of the diagnostic measurements and multiple cycles as considered in \cite{Son2021} affects the TSM heat engine proposed here. Finally, while we have shown that the coherent average work can in some situations exceed the TMA work, its relevance in terms of whether one can out-couple and extract this work in a realistic setting remains an outstanding problem as well \cite{Thingna2019}. 

\vspace{0.1in}

\section{Acknowledgement}
This work was supported and enabled by the Excellence-in-Research Fellowship of IIT Gandhinagar (B.~P.~V) and the  Department of Science \& Technology Science and Engineering Research Board (India) Start-up Research Grant No. SRG/2019/001585 (B.~P.~V \& M.~S.~A) . We thank Chinmayee Mishra for help with making the schematic in Figure. \eqref{fig:figure1}. M.~S.~A thanks Rahul Shastri and Jayanth Jayakumar for many fruitful discussions.

\bibliography{references}

\begin{thebibliography}{39}%
\makeatletter
\providecommand \@ifxundefined [1]{%
 \@ifx{#1\undefined}
}%
\providecommand \@ifnum [1]{%
 \ifnum #1\expandafter \@firstoftwo
 \else \expandafter \@secondoftwo
 \fi
}%
\providecommand \@ifx [1]{%
 \ifx #1\expandafter \@firstoftwo
 \else \expandafter \@secondoftwo
 \fi
}%
\providecommand \natexlab [1]{#1}%
\providecommand \enquote  [1]{``#1''}%
\providecommand \bibnamefont  [1]{#1}%
\providecommand \bibfnamefont [1]{#1}%
\providecommand \citenamefont [1]{#1}%
\providecommand \href@noop [0]{\@secondoftwo}%
\providecommand \href [0]{\begingroup \@sanitize@url \@href}%
\providecommand \@href[1]{\@@startlink{#1}\@@href}%
\providecommand \@@href[1]{\endgroup#1\@@endlink}%
\providecommand \@sanitize@url [0]{\catcode `\\12\catcode `\$12\catcode
  `\&12\catcode `\#12\catcode `\^12\catcode `\_12\catcode `\%12\relax}%
\providecommand \@@startlink[1]{}%
\providecommand \@@endlink[0]{}%
\providecommand \url  [0]{\begingroup\@sanitize@url \@url }%
\providecommand \@url [1]{\endgroup\@href {#1}{\urlprefix }}%
\providecommand \urlprefix  [0]{URL }%
\providecommand \Eprint [0]{\href }%
\providecommand \doibase [0]{http://dx.doi.org/}%
\providecommand \selectlanguage [0]{\@gobble}%
\providecommand \bibinfo  [0]{\@secondoftwo}%
\providecommand \bibfield  [0]{\@secondoftwo}%
\providecommand \translation [1]{[#1]}%
\providecommand \BibitemOpen [0]{}%
\providecommand \bibitemStop [0]{}%
\providecommand \bibitemNoStop [0]{.\EOS\space}%
\providecommand \EOS [0]{\spacefactor3000\relax}%
\providecommand \BibitemShut  [1]{\csname bibitem#1\endcsname}%
\let\auto@bib@innerbib\@empty
\bibitem [{\citenamefont {Vinjanampathy}\ and\ \citenamefont
  {Anders}(2016)}]{Vinjanampathy2016}%
  \BibitemOpen
  \bibfield  {author} {\bibinfo {author} {\bibfnamefont {S.}~\bibnamefont
  {Vinjanampathy}}\ and\ \bibinfo {author} {\bibfnamefont {J.}~\bibnamefont
  {Anders}},\ }\href {\doibase 10.1080/00107514.2016.1201896} {\bibfield
  {journal} {\bibinfo  {journal} {Contemporary Physics}\ }\textbf {\bibinfo
  {volume} {57}},\ \bibinfo {pages} {545} (\bibinfo {year} {2016})}\BibitemShut
  {NoStop}%
\bibitem [{\citenamefont {Alicki}\ and\ \citenamefont
  {Kosloff}(2018)}]{Alicki2018}%
  \BibitemOpen
  \bibfield  {author} {\bibinfo {author} {\bibfnamefont {R.}~\bibnamefont
  {Alicki}}\ and\ \bibinfo {author} {\bibfnamefont {R.}~\bibnamefont
  {Kosloff}},\ }in\ \href {\doibase 10.1007/978-3-319-99046-0_1} {\emph
  {\bibinfo {booktitle} {Thermodynamics in the Quantum Regime, Fundamental
  Theories of Physics}}},\ Vol.\ \bibinfo {volume} {195},\ \bibinfo {editor}
  {edited by\ \bibinfo {editor} {\bibfnamefont {F.}~\bibnamefont {Binder}},
  \bibinfo {editor} {\bibfnamefont {L.}~\bibnamefont {Correa}}, \bibinfo
  {editor} {\bibfnamefont {C.}~\bibnamefont {Gogolin}}, \bibinfo {editor}
  {\bibfnamefont {J.}~\bibnamefont {Anders}}, \ and\ \bibinfo {editor}
  {\bibfnamefont {G.}~\bibnamefont {Adesso}}}\ (\bibinfo  {publisher}
  {Springer},\ \bibinfo {year} {2018})\ Chap.~\bibinfo {chapter} {1}, pp.\
  \bibinfo {pages} {1--33}\BibitemShut {NoStop}%
\bibitem [{\citenamefont {Levy}\ and\ \citenamefont
  {Gelbwaser-Klimovsky}(2018)}]{Levy2018}%
  \BibitemOpen
  \bibfield  {author} {\bibinfo {author} {\bibfnamefont {A.}~\bibnamefont
  {Levy}}\ and\ \bibinfo {author} {\bibfnamefont {D.}~\bibnamefont
  {Gelbwaser-Klimovsky}},\ }in\ \href {\doibase 10.1007/978-3-319-99046-0_4}
  {\emph {\bibinfo {booktitle} {Thermodynamics in the Quantum Regime,
  Fundamental Theories of Physics}}},\ Vol.\ \bibinfo {volume} {195},\ \bibinfo
  {editor} {edited by\ \bibinfo {editor} {\bibfnamefont {F.}~\bibnamefont
  {Binder}}, \bibinfo {editor} {\bibfnamefont {L.}~\bibnamefont {Correa}},
  \bibinfo {editor} {\bibfnamefont {C.}~\bibnamefont {Gogolin}}, \bibinfo
  {editor} {\bibfnamefont {J.}~\bibnamefont {Anders}}, \ and\ \bibinfo {editor}
  {\bibfnamefont {G.}~\bibnamefont {Adesso}}}\ (\bibinfo  {publisher}
  {Springer},\ \bibinfo {year} {2018})\ Chap.~\bibinfo {chapter} {4}, pp.\
  \bibinfo {pages} {87--126}\BibitemShut {NoStop}%
\bibitem [{\citenamefont {Kammerlander}\ and\ \citenamefont
  {Anders}(2016)}]{Kammerlander2016}%
  \BibitemOpen
  \bibfield  {author} {\bibinfo {author} {\bibfnamefont {P.}~\bibnamefont
  {Kammerlander}}\ and\ \bibinfo {author} {\bibfnamefont {J.}~\bibnamefont
  {Anders}},\ }\href {\doibase 10.1038/srep22174} {\bibfield  {journal}
  {\bibinfo  {journal} {Scientific Reports}\ }\textbf {\bibinfo {volume} {6}},\
  \bibinfo {pages} {22174} (\bibinfo {year} {2016})}\BibitemShut {NoStop}%
\bibitem [{\citenamefont {Ro{\ss}nagel}\ \emph {et~al.}(2014)\citenamefont
  {Ro{\ss}nagel}, \citenamefont {Abah}, \citenamefont {Schmidt-Kaler},
  \citenamefont {Singer},\ and\ \citenamefont {Lutz}}]{Rossnagel2014}%
  \BibitemOpen
  \bibfield  {author} {\bibinfo {author} {\bibfnamefont {J.}~\bibnamefont
  {Ro{\ss}nagel}}, \bibinfo {author} {\bibfnamefont {O.}~\bibnamefont {Abah}},
  \bibinfo {author} {\bibfnamefont {F.}~\bibnamefont {Schmidt-Kaler}}, \bibinfo
  {author} {\bibfnamefont {K.}~\bibnamefont {Singer}}, \ and\ \bibinfo {author}
  {\bibfnamefont {E.}~\bibnamefont {Lutz}},\ }\href {\doibase
  10.1103/PhysRevLett.112.030602} {\bibfield  {journal} {\bibinfo  {journal}
  {Physical Review Letters}\ }\textbf {\bibinfo {volume} {112}},\ \bibinfo
  {pages} {030602} (\bibinfo {year} {2014})}\BibitemShut {NoStop}%
\bibitem [{\citenamefont {Niedenzu}\ \emph {et~al.}(2016)\citenamefont
  {Niedenzu}, \citenamefont {Gelbwaser-Klimovsky}, \citenamefont {Kofman},\
  and\ \citenamefont {Kurizki}}]{Niedenzu2016}%
  \BibitemOpen
  \bibfield  {author} {\bibinfo {author} {\bibfnamefont {W.}~\bibnamefont
  {Niedenzu}}, \bibinfo {author} {\bibfnamefont {D.}~\bibnamefont
  {Gelbwaser-Klimovsky}}, \bibinfo {author} {\bibfnamefont {A.~G.}\
  \bibnamefont {Kofman}}, \ and\ \bibinfo {author} {\bibfnamefont
  {G.}~\bibnamefont {Kurizki}},\ }\href {\doibase
  10.1088/1367-2630/18/8/083012} {\bibfield  {journal} {\bibinfo  {journal}
  {New Journal of Physics}\ }\textbf {\bibinfo {volume} {18}},\ \bibinfo
  {pages} {083012} (\bibinfo {year} {2016})}\BibitemShut {NoStop}%
\bibitem [{\citenamefont {Pozas-Kerstjens}\ \emph {et~al.}(2018)\citenamefont
  {Pozas-Kerstjens}, \citenamefont {Brown},\ and\ \citenamefont
  {Hovhannisyan}}]{Pozas2018}%
  \BibitemOpen
  \bibfield  {author} {\bibinfo {author} {\bibfnamefont {A.}~\bibnamefont
  {Pozas-Kerstjens}}, \bibinfo {author} {\bibfnamefont {E.~G.}\ \bibnamefont
  {Brown}}, \ and\ \bibinfo {author} {\bibfnamefont {K.~V.}\ \bibnamefont
  {Hovhannisyan}},\ }\href {\doibase 10.1088/1367-2630/AABA02} {\bibfield
  {journal} {\bibinfo  {journal} {New Journal of Physics}\ }\textbf {\bibinfo
  {volume} {20}},\ \bibinfo {pages} {043034} (\bibinfo {year}
  {2018})}\BibitemShut {NoStop}%
\bibitem [{\citenamefont {Jacobs}(2012)}]{Jacobs2012}%
  \BibitemOpen
  \bibfield  {author} {\bibinfo {author} {\bibfnamefont {K.}~\bibnamefont
  {Jacobs}},\ }\href {\doibase 10.1103/PhysRevE.86.040106} {\bibfield
  {journal} {\bibinfo  {journal} {Physical Review E}\ }\textbf {\bibinfo
  {volume} {86}},\ \bibinfo {pages} {040106} (\bibinfo {year}
  {2012})}\BibitemShut {NoStop}%
\bibitem [{\citenamefont {Yi}\ \emph {et~al.}(2017)\citenamefont {Yi},
  \citenamefont {Talkner},\ and\ \citenamefont {Kim}}]{Yi2017}%
  \BibitemOpen
  \bibfield  {author} {\bibinfo {author} {\bibfnamefont {J.}~\bibnamefont
  {Yi}}, \bibinfo {author} {\bibfnamefont {P.}~\bibnamefont {Talkner}}, \ and\
  \bibinfo {author} {\bibfnamefont {Y.~W.}\ \bibnamefont {Kim}},\ }\href
  {\doibase 10.1103/PhysRevE.96.022108} {\bibfield  {journal} {\bibinfo
  {journal} {Physical Review E}\ }\textbf {\bibinfo {volume} {96}},\ \bibinfo
  {pages} {022108} (\bibinfo {year} {2017})}\BibitemShut {NoStop}%
\bibitem [{\citenamefont {Jordan}\ \emph {et~al.}(2020)\citenamefont {Jordan},
  \citenamefont {Elouard},\ and\ \citenamefont {Auff{\`{e}}ves}}]{Jordan2020}%
  \BibitemOpen
  \bibfield  {author} {\bibinfo {author} {\bibfnamefont {A.~N.}\ \bibnamefont
  {Jordan}}, \bibinfo {author} {\bibfnamefont {C.}~\bibnamefont {Elouard}}, \
  and\ \bibinfo {author} {\bibfnamefont {A.}~\bibnamefont {Auff{\`{e}}ves}},\
  }\href {\doibase 10.1007/s40509-019-00217-2} {\bibfield  {journal} {\bibinfo
  {journal} {Quantum Studies: Mathematics and Foundations}\ }\textbf {\bibinfo
  {volume} {7}},\ \bibinfo {pages} {203} (\bibinfo {year} {2020})}\BibitemShut
  {NoStop}%
\bibitem [{\citenamefont {Ding}\ \emph {et~al.}(2018)\citenamefont {Ding},
  \citenamefont {Yi}, \citenamefont {Kim},\ and\ \citenamefont
  {Talkner}}]{Ding2018a}%
  \BibitemOpen
  \bibfield  {author} {\bibinfo {author} {\bibfnamefont {X.}~\bibnamefont
  {Ding}}, \bibinfo {author} {\bibfnamefont {J.}~\bibnamefont {Yi}}, \bibinfo
  {author} {\bibfnamefont {Y.~W.}\ \bibnamefont {Kim}}, \ and\ \bibinfo
  {author} {\bibfnamefont {P.}~\bibnamefont {Talkner}},\ }\href {\doibase
  10.1103/PhysRevE.98.042122} {\bibfield  {journal} {\bibinfo  {journal}
  {Physical Review E}\ }\textbf {\bibinfo {volume} {98}},\ \bibinfo {pages}
  {042122} (\bibinfo {year} {2018})}\BibitemShut {NoStop}%
\bibitem [{\citenamefont {Anka}\ \emph {et~al.}(2021)\citenamefont {Anka},
  \citenamefont {de~Oliveira},\ and\ \citenamefont {Jonathan}}]{Anka2021}%
  \BibitemOpen
  \bibfield  {author} {\bibinfo {author} {\bibfnamefont {M.}~\bibnamefont
  {Anka}}, \bibinfo {author} {\bibfnamefont {T.~R.}\ \bibnamefont
  {de~Oliveira}}, \ and\ \bibinfo {author} {\bibfnamefont {D.}~\bibnamefont
  {Jonathan}},\ }\href {https://arxiv.org/abs/2108.02229} {\bibfield  {journal}
  {\bibinfo  {journal} {arXiv:2108.02229}\ } (\bibinfo {year}
  {2021})}\BibitemShut {NoStop}%
\bibitem [{\citenamefont {Lin}\ \emph {et~al.}(2021)\citenamefont {Lin},
  \citenamefont {Su}, \citenamefont {Chen}, \citenamefont {Chen},\ and\
  \citenamefont {Santos}}]{Lin2021}%
  \BibitemOpen
  \bibfield  {author} {\bibinfo {author} {\bibfnamefont {Z.}~\bibnamefont
  {Lin}}, \bibinfo {author} {\bibfnamefont {S.}~\bibnamefont {Su}}, \bibinfo
  {author} {\bibfnamefont {J.}~\bibnamefont {Chen}}, \bibinfo {author}
  {\bibfnamefont {J.}~\bibnamefont {Chen}}, \ and\ \bibinfo {author}
  {\bibfnamefont {J.~F.~G.}\ \bibnamefont {Santos}},\ }\href
  {http://arxiv.org/abs/2108.07995} {\bibfield  {journal} {\bibinfo  {journal}
  {arXiv:2108.07995}\ } (\bibinfo {year} {\hspace{-0.05in} 2021})}\BibitemShut
  {NoStop}%
\bibitem [{\citenamefont {Behzadi}(2021)}]{Behzadi2021}%
  \BibitemOpen
  \bibfield  {author} {\bibinfo {author} {\bibfnamefont {N.}~\bibnamefont
  {Behzadi}},\ }\href {\doibase 10.1088/1751-8121/abca74} {\bibfield  {journal}
  {\bibinfo  {journal} {Journal of Physics A: Mathematical and Theoretical}\
  }\textbf {\bibinfo {volume} {54}},\ \bibinfo {pages} {015304} (\bibinfo
  {year} {2021})}\BibitemShut {NoStop}%
\bibitem [{\citenamefont {Su}\ \emph {et~al.}(2021)\citenamefont {Su},
  \citenamefont {Lin},\ and\ \citenamefont {Chen}}]{Su2021}%
  \BibitemOpen
  \bibfield  {author} {\bibinfo {author} {\bibfnamefont {S.}~\bibnamefont
  {Su}}, \bibinfo {author} {\bibfnamefont {Z.}~\bibnamefont {Lin}}, \ and\
  \bibinfo {author} {\bibfnamefont {J.}~\bibnamefont {Chen}},\ }\href
  {http://arxiv.org/abs/2109.10796} {\bibfield  {journal} {\bibinfo  {journal}
  {arXiv:2109.10796}\ } (\bibinfo {year} {2021})}\BibitemShut {NoStop}%
\bibitem [{\citenamefont {Elouard}\ and\ \citenamefont
  {Jordan}(2018)}]{Elouard2018}%
  \BibitemOpen
  \bibfield  {author} {\bibinfo {author} {\bibfnamefont {C.}~\bibnamefont
  {Elouard}}\ and\ \bibinfo {author} {\bibfnamefont {A.~N.}\ \bibnamefont
  {Jordan}},\ }\href {\doibase 10.1103/PhysRevLett.120.260601} {\bibfield
  {journal} {\bibinfo  {journal} {Physical Review Letters}\ }\textbf {\bibinfo
  {volume} {120}},\ \bibinfo {pages} {260601} (\bibinfo {year}
  {2018})}\BibitemShut {NoStop}%
\bibitem [{\citenamefont {Buffoni}\ \emph {et~al.}(2019)\citenamefont
  {Buffoni}, \citenamefont {Solfanelli}, \citenamefont {Verrucchi},
  \citenamefont {Cuccoli},\ and\ \citenamefont {Campisi}}]{Buffoni2019}%
  \BibitemOpen
  \bibfield  {author} {\bibinfo {author} {\bibfnamefont {L.}~\bibnamefont
  {Buffoni}}, \bibinfo {author} {\bibfnamefont {A.}~\bibnamefont {Solfanelli}},
  \bibinfo {author} {\bibfnamefont {P.}~\bibnamefont {Verrucchi}}, \bibinfo
  {author} {\bibfnamefont {A.}~\bibnamefont {Cuccoli}}, \ and\ \bibinfo
  {author} {\bibfnamefont {M.}~\bibnamefont {Campisi}},\ }\href {\doibase
  10.1103/PhysRevLett.122.070603} {\bibfield  {journal} {\bibinfo  {journal}
  {Physical Review Letters}\ }\textbf {\bibinfo {volume} {122}},\ \bibinfo
  {pages} {070603} (\bibinfo {year} {2019})}\BibitemShut {NoStop}%
\bibitem [{\citenamefont {Bresque}\ \emph {et~al.}(2021)\citenamefont
  {Bresque}, \citenamefont {Camati}, \citenamefont {Rogers}, \citenamefont
  {Murch}, \citenamefont {Jordan},\ and\ \citenamefont
  {Auff{\`{e}}ves}}]{Bresque2021}%
  \BibitemOpen
  \bibfield  {author} {\bibinfo {author} {\bibfnamefont {L.}~\bibnamefont
  {Bresque}}, \bibinfo {author} {\bibfnamefont {P.~A.}\ \bibnamefont {Camati}},
  \bibinfo {author} {\bibfnamefont {S.}~\bibnamefont {Rogers}}, \bibinfo
  {author} {\bibfnamefont {K.}~\bibnamefont {Murch}}, \bibinfo {author}
  {\bibfnamefont {A.~N.}\ \bibnamefont {Jordan}}, \ and\ \bibinfo {author}
  {\bibfnamefont {A.}~\bibnamefont {Auff{\`{e}}ves}},\ }\href {\doibase
  10.1103/PhysRevLett.126.120605} {\bibfield  {journal} {\bibinfo  {journal}
  {Physical Review Letters}\ }\textbf {\bibinfo {volume} {126}},\ \bibinfo
  {pages} {120605} (\bibinfo {year} {2021})}\BibitemShut {NoStop}%
\bibitem [{\citenamefont {Opatrn{\'{y}}}\ \emph {et~al.}(2021)\citenamefont
  {Opatrn{\'{y}}}, \citenamefont {Misra},\ and\ \citenamefont
  {Kurizki}}]{Opatrny2021}%
  \BibitemOpen
  \bibfield  {author} {\bibinfo {author} {\bibfnamefont {T.}~\bibnamefont
  {Opatrn{\'{y}}}}, \bibinfo {author} {\bibfnamefont {A.}~\bibnamefont
  {Misra}}, \ and\ \bibinfo {author} {\bibfnamefont {G.}~\bibnamefont
  {Kurizki}},\ }\href {\doibase 10.1103/PhysRevLett.127.040602} {\bibfield
  {journal} {\bibinfo  {journal} {Physical Review Letters}\ }\textbf {\bibinfo
  {volume} {127}},\ \bibinfo {pages} {040602} (\bibinfo {year}
  {2021})}\BibitemShut {NoStop}%
\bibitem [{\citenamefont {Quan}\ \emph {et~al.}(2007)\citenamefont {Quan},
  \citenamefont {Liu}, \citenamefont {Sun},\ and\ \citenamefont
  {Nori}}]{Quan2007}%
  \BibitemOpen
  \bibfield  {author} {\bibinfo {author} {\bibfnamefont {H.~T.}\ \bibnamefont
  {Quan}}, \bibinfo {author} {\bibfnamefont {Y.-x.}\ \bibnamefont {Liu}},
  \bibinfo {author} {\bibfnamefont {C.~P.}\ \bibnamefont {Sun}}, \ and\
  \bibinfo {author} {\bibfnamefont {F.}~\bibnamefont {Nori}},\ }\href {\doibase
  10.1103/PhysRevE.76.031105} {\bibfield  {journal} {\bibinfo  {journal}
  {Physical Review E}\ }\textbf {\bibinfo {volume} {76}},\ \bibinfo {pages}
  {031105} (\bibinfo {year} {2007})}\BibitemShut {NoStop}%
\bibitem [{\citenamefont {Campisi}\ \emph {et~al.}(2015)\citenamefont
  {Campisi}, \citenamefont {Pekola},\ and\ \citenamefont
  {Fazio}}]{Campisi2015}%
  \BibitemOpen
  \bibfield  {author} {\bibinfo {author} {\bibfnamefont {M.}~\bibnamefont
  {Campisi}}, \bibinfo {author} {\bibfnamefont {J.}~\bibnamefont {Pekola}}, \
  and\ \bibinfo {author} {\bibfnamefont {R.}~\bibnamefont {Fazio}},\ }\href
  {\doibase 10.1088/1367-2630/17/3/035012} {\bibfield  {journal} {\bibinfo
  {journal} {New Journal of Physics}\ }\textbf {\bibinfo {volume} {17}},\
  \bibinfo {pages} {1} (\bibinfo {year} {2015})}\BibitemShut {NoStop}%
\bibitem [{\citenamefont {Sacchi}(2021{\natexlab{a}})}]{Sacchi2021a}%
  \BibitemOpen
  \bibfield  {author} {\bibinfo {author} {\bibfnamefont {M.~F.}\ \bibnamefont
  {Sacchi}},\ }\href {\doibase 10.1103/physreva.104.012217} {\bibfield
  {journal} {\bibinfo  {journal} {Physical Review A}\ }\textbf {\bibinfo
  {volume} {104}},\ \bibinfo {pages} {012217} (\bibinfo {year}
  {2021}{\natexlab{a}})}\BibitemShut {NoStop}%
\bibitem [{\citenamefont {Sacchi}(2021{\natexlab{b}})}]{Sacchi2021b}%
  \BibitemOpen
  \bibfield  {author} {\bibinfo {author} {\bibfnamefont {M.~F.}\ \bibnamefont
  {Sacchi}},\ }\href {\doibase 10.1103/PhysRevE.103.012111} {\bibfield
  {journal} {\bibinfo  {journal} {Physical Review E}\ }\textbf {\bibinfo
  {volume} {103}},\ \bibinfo {pages} {012111} (\bibinfo {year}
  {2021}{\natexlab{b}})}\BibitemShut {NoStop}%
\bibitem [{\citenamefont {Son}\ \emph {et~al.}(2021)\citenamefont {Son},
  \citenamefont {Talkner},\ and\ \citenamefont {Thingna}}]{Son2021}%
  \BibitemOpen
  \bibfield  {author} {\bibinfo {author} {\bibfnamefont {J.}~\bibnamefont
  {Son}}, \bibinfo {author} {\bibfnamefont {P.}~\bibnamefont {Talkner}}, \ and\
  \bibinfo {author} {\bibfnamefont {J.}~\bibnamefont {Thingna}},\ }\href
  {\doibase 10.1103/PRXQuantum.2.040328} {\bibfield  {journal} {\bibinfo
  {journal} {PRX Quantum}\ }\textbf {\bibinfo {volume} {2}},\ \bibinfo {pages}
  {040328} (\bibinfo {year} {2021})}\BibitemShut {NoStop}%
\bibitem [{\citenamefont {Solinas}\ and\ \citenamefont
  {Gasparinetti}(2015)}]{Solinas2015}%
  \BibitemOpen
  \bibfield  {author} {\bibinfo {author} {\bibfnamefont {P.}~\bibnamefont
  {Solinas}}\ and\ \bibinfo {author} {\bibfnamefont {S.}~\bibnamefont
  {Gasparinetti}},\ }\href {\doibase 10.1103/PhysRevE.92.042150} {\bibfield
  {journal} {\bibinfo  {journal} {Physical Review E}\ }\textbf {\bibinfo
  {volume} {92}},\ \bibinfo {pages} {042150} (\bibinfo {year}
  {2015})}\BibitemShut {NoStop}%
\bibitem [{\citenamefont {Talkner}\ and\ \citenamefont
  {H{\"{a}}nggi}(2016)}]{Talkner2016}%
  \BibitemOpen
  \bibfield  {author} {\bibinfo {author} {\bibfnamefont {P.}~\bibnamefont
  {Talkner}}\ and\ \bibinfo {author} {\bibfnamefont {P.}~\bibnamefont
  {H{\"{a}}nggi}},\ }\href {\doibase 10.1103/PhysRevE.93.022131} {\bibfield
  {journal} {\bibinfo  {journal} {Physical Review E}\ }\textbf {\bibinfo
  {volume} {93}},\ \bibinfo {pages} {022131} (\bibinfo {year}
  {2016})}\BibitemShut {NoStop}%
\bibitem [{\citenamefont {Perarnau-Llobet}\ \emph {et~al.}(2017)\citenamefont
  {Perarnau-Llobet}, \citenamefont {B{\"{a}}umer}, \citenamefont
  {Hovhannisyan}, \citenamefont {Huber},\ and\ \citenamefont
  {Acin}}]{Perarnau-Llobet2017}%
  \BibitemOpen
  \bibfield  {author} {\bibinfo {author} {\bibfnamefont {M.}~\bibnamefont
  {Perarnau-Llobet}}, \bibinfo {author} {\bibfnamefont {E.}~\bibnamefont
  {B{\"{a}}umer}}, \bibinfo {author} {\bibfnamefont {K.~V.}\ \bibnamefont
  {Hovhannisyan}}, \bibinfo {author} {\bibfnamefont {M.}~\bibnamefont {Huber}},
  \ and\ \bibinfo {author} {\bibfnamefont {A.}~\bibnamefont {Acin}},\ }\href
  {\doibase 10.1103/PhysRevLett.118.070601} {\bibfield  {journal} {\bibinfo
  {journal} {Physical Review Letters}\ }\textbf {\bibinfo {volume} {118}},\
  \bibinfo {pages} {070601} (\bibinfo {year} {2017})}\BibitemShut {NoStop}%
\bibitem [{\citenamefont {Rezek}\ and\ \citenamefont
  {Kosloff}(2006)}]{Rezek2006}%
  \BibitemOpen
  \bibfield  {author} {\bibinfo {author} {\bibfnamefont {Y.}~\bibnamefont
  {Rezek}}\ and\ \bibinfo {author} {\bibfnamefont {R.}~\bibnamefont
  {Kosloff}},\ }\href {\doibase 10.1088/1367-2630/8/5/083} {\bibfield
  {journal} {\bibinfo  {journal} {New Journal of Physics}\ }\textbf {\bibinfo
  {volume} {8}},\ \bibinfo {pages} {83} (\bibinfo {year} {2006})}\BibitemShut
  {NoStop}%
\bibitem [{\citenamefont {Dann}\ \emph {et~al.}(2020)\citenamefont {Dann},
  \citenamefont {Kosloff},\ and\ \citenamefont {Salamon}}]{Dann2020}%
  \BibitemOpen
  \bibfield  {author} {\bibinfo {author} {\bibfnamefont {R.}~\bibnamefont
  {Dann}}, \bibinfo {author} {\bibfnamefont {R.}~\bibnamefont {Kosloff}}, \
  and\ \bibinfo {author} {\bibfnamefont {P.}~\bibnamefont {Salamon}},\ }\href
  {\doibase 10.3390/e22111255} {\bibfield  {journal} {\bibinfo  {journal}
  {Entropy}\ }\textbf {\bibinfo {volume} {22}},\ \bibinfo {pages} {1255}
  (\bibinfo {year} {2020})}\BibitemShut {NoStop}%
\bibitem [{Note1()}]{Note1}%
  \BibitemOpen
  \bibinfo {note} {Note that relations such as $Q_c = \Delta E_A - w$,
  $E_{\protect \mathcal {M}} = -\Delta E_A$ at the level of the stochastic
  variables are not justified in general.}\BibitemShut {Stop}%
\bibitem [{\citenamefont {Barato}\ and\ \citenamefont
  {Seifert}(2015)}]{Barato2015}%
  \BibitemOpen
  \bibfield  {author} {\bibinfo {author} {\bibfnamefont {A.~C.}\ \bibnamefont
  {Barato}}\ and\ \bibinfo {author} {\bibfnamefont {U.}~\bibnamefont
  {Seifert}},\ }\href {\doibase 10.1103/PhysRevLett.114.158101} {\bibfield
  {journal} {\bibinfo  {journal} {Physical Review Letters}\ }\textbf {\bibinfo
  {volume} {114}},\ \bibinfo {pages} {158101} (\bibinfo {year}
  {2015})}\BibitemShut {NoStop}%
\bibitem [{\citenamefont {Timpanaro}\ \emph {et~al.}(2019)\citenamefont
  {Timpanaro}, \citenamefont {Guarnieri}, \citenamefont {Goold},\ and\
  \citenamefont {Landi}}]{Timpanaro2019}%
  \BibitemOpen
  \bibfield  {author} {\bibinfo {author} {\bibfnamefont {A.~M.}\ \bibnamefont
  {Timpanaro}}, \bibinfo {author} {\bibfnamefont {G.}~\bibnamefont
  {Guarnieri}}, \bibinfo {author} {\bibfnamefont {J.}~\bibnamefont {Goold}}, \
  and\ \bibinfo {author} {\bibfnamefont {G.~T.}\ \bibnamefont {Landi}},\ }\href
  {\doibase 10.1103/PhysRevLett.123.090604} {\bibfield  {journal} {\bibinfo
  {journal} {Physical Review Letters}\ }\textbf {\bibinfo {volume} {123}},\
  \bibinfo {pages} {090604} (\bibinfo {year} {2019})}\BibitemShut {NoStop}%
\bibitem [{\citenamefont {Pekola}(2015)}]{Pekola2015}%
  \BibitemOpen
  \bibfield  {author} {\bibinfo {author} {\bibfnamefont {J.~P.}\ \bibnamefont
  {Pekola}},\ }\href {\doibase 10.1038/nphys3169} {\bibfield  {journal}
  {\bibinfo  {journal} {Nature Physics 2014 11:2}\ }\textbf {\bibinfo {volume}
  {11}},\ \bibinfo {pages} {118} (\bibinfo {year} {2015})}\BibitemShut
  {NoStop}%
\bibitem [{\citenamefont {Pekola}\ and\ \citenamefont
  {Khaymovich}(2019)}]{Pekola2019}%
  \BibitemOpen
  \bibfield  {author} {\bibinfo {author} {\bibfnamefont {J.~P.}\ \bibnamefont
  {Pekola}}\ and\ \bibinfo {author} {\bibfnamefont {I.~M.}\ \bibnamefont
  {Khaymovich}},\ }\href {\doibase 10.1146/annurev-conmatphys-033117-054120}
  {\bibfield  {journal} {\bibinfo  {journal} {Annual Review of Condensed Matter
  Physics}\ }\textbf {\bibinfo {volume} {10}},\ \bibinfo {pages} {193}
  (\bibinfo {year} {2019})}\BibitemShut {NoStop}%
\bibitem [{\citenamefont {Murch}\ \emph {et~al.}(2013)\citenamefont {Murch},
  \citenamefont {Weber}, \citenamefont {Macklin},\ and\ \citenamefont
  {Siddiqi}}]{Murch2013}%
  \BibitemOpen
  \bibfield  {author} {\bibinfo {author} {\bibfnamefont {K.~W.}\ \bibnamefont
  {Murch}}, \bibinfo {author} {\bibfnamefont {S.~J.}\ \bibnamefont {Weber}},
  \bibinfo {author} {\bibfnamefont {C.}~\bibnamefont {Macklin}}, \ and\
  \bibinfo {author} {\bibfnamefont {I.}~\bibnamefont {Siddiqi}},\ }\href
  {\doibase 10.1038/nature12539} {\bibfield  {journal} {\bibinfo  {journal}
  {Nature}\ }\textbf {\bibinfo {volume} {502}},\ \bibinfo {pages} {211}
  (\bibinfo {year} {2013})}\BibitemShut {NoStop}%
\bibitem [{\citenamefont {Weber}\ \emph {et~al.}(2014)\citenamefont {Weber},
  \citenamefont {Chantasri}, \citenamefont {Dressel}, \citenamefont {Jordan},
  \citenamefont {Murch},\ and\ \citenamefont {Siddiqi}}]{Weber2014}%
  \BibitemOpen
  \bibfield  {author} {\bibinfo {author} {\bibfnamefont {S.~J.}\ \bibnamefont
  {Weber}}, \bibinfo {author} {\bibfnamefont {A.}~\bibnamefont {Chantasri}},
  \bibinfo {author} {\bibfnamefont {J.}~\bibnamefont {Dressel}}, \bibinfo
  {author} {\bibfnamefont {A.~N.}\ \bibnamefont {Jordan}}, \bibinfo {author}
  {\bibfnamefont {K.~W.}\ \bibnamefont {Murch}}, \ and\ \bibinfo {author}
  {\bibfnamefont {I.}~\bibnamefont {Siddiqi}},\ }\href {\doibase
  10.1038/nature13559} {\bibfield  {journal} {\bibinfo  {journal} {Nature}\
  }\textbf {\bibinfo {volume} {511}},\ \bibinfo {pages} {570} (\bibinfo {year}
  {2014})}\BibitemShut {NoStop}%
\bibitem [{\citenamefont {Murch}\ \emph {et~al.}(2016)\citenamefont {Murch},
  \citenamefont {Vijay},\ and\ \citenamefont {Siddiqi}}]{Murch2016}%
  \BibitemOpen
  \bibfield  {author} {\bibinfo {author} {\bibfnamefont {K.~W.}\ \bibnamefont
  {Murch}}, \bibinfo {author} {\bibfnamefont {R.}~\bibnamefont {Vijay}}, \ and\
  \bibinfo {author} {\bibfnamefont {I.}~\bibnamefont {Siddiqi}},\ }in\ \href
  {\doibase 10.1007/978-3-319-24091-6_7} {\emph {\bibinfo {booktitle}
  {Superconducting Devices in Quantum Optics}}},\ \bibinfo {editor} {edited by\
  \bibinfo {editor} {\bibfnamefont {R.~H.}\ \bibnamefont {Hadfield}}\ and\
  \bibinfo {editor} {\bibfnamefont {G.}~\bibnamefont {Johansson}}}\ (\bibinfo
  {publisher} {Springer},\ \bibinfo {year} {2016})\ Chap.~\bibinfo {chapter}
  {7}, pp.\ \bibinfo {pages} {163--185}\BibitemShut {NoStop}%
\bibitem [{\citenamefont {Naghiloo}\ \emph {et~al.}(2020)\citenamefont
  {Naghiloo}, \citenamefont {Tan}, \citenamefont {Harrington}, \citenamefont
  {Alonso}, \citenamefont {Lutz}, \citenamefont {Romito},\ and\ \citenamefont
  {Murch}}]{Naghiloo2020}%
  \BibitemOpen
  \bibfield  {author} {\bibinfo {author} {\bibfnamefont {M.}~\bibnamefont
  {Naghiloo}}, \bibinfo {author} {\bibfnamefont {D.}~\bibnamefont {Tan}},
  \bibinfo {author} {\bibfnamefont {P.~M.}\ \bibnamefont {Harrington}},
  \bibinfo {author} {\bibfnamefont {J.~J.}\ \bibnamefont {Alonso}}, \bibinfo
  {author} {\bibfnamefont {E.}~\bibnamefont {Lutz}}, \bibinfo {author}
  {\bibfnamefont {A.}~\bibnamefont {Romito}}, \ and\ \bibinfo {author}
  {\bibfnamefont {K.~W.}\ \bibnamefont {Murch}},\ }\href {\doibase
  10.1103/PhysRevLett.124.110604} {\bibfield  {journal} {\bibinfo  {journal}
  {Physical Review Letters}\ }\textbf {\bibinfo {volume} {124}},\ \bibinfo
  {pages} {110604} (\bibinfo {year} {2020})}\BibitemShut {NoStop}%
\bibitem [{\citenamefont {Thingna}\ \emph {et~al.}(2019)\citenamefont
  {Thingna}, \citenamefont {Esposito},\ and\ \citenamefont
  {Barra}}]{Thingna2019}%
  \BibitemOpen
  \bibfield  {author} {\bibinfo {author} {\bibfnamefont {J.}~\bibnamefont
  {Thingna}}, \bibinfo {author} {\bibfnamefont {M.}~\bibnamefont {Esposito}}, \
  and\ \bibinfo {author} {\bibfnamefont {F.}~\bibnamefont {Barra}},\ }\href
  {\doibase 10.1103/PhysRevE.99.042142} {\bibfield  {journal} {\bibinfo
  {journal} {Physical Review E}\ }\textbf {\bibinfo {volume} {99}},\ \bibinfo
  {pages} {042142} (\bibinfo {year} {2019})}\BibitemShut {NoStop}%
\end{thebibliography}%

\end{document}